\documentclass{fac}

\usepackage{etex}
\usepackage{amsmath}
\allowdisplaybreaks
\usepackage{mathtools}
\usepackage{txfonts}
\usepackage{fancyvrb}
\usepackage{hyperref}
\usepackage{listings}
\usepackage{paralist}
\usepackage{multirow}
\usepackage{marvosym}
\usepackage[nounderscore]{syntax}
\usepackage{pstricks,pst-node,pst-tree,pstricks-add}  
\usepackage{tikz}  
\usetikzlibrary{arrows, shapes, shadows, calc, decorations.markings}
\tikzset{
  ->-/.style={
    thick,
    postaction = {decorate},
    decoration = {
      markings,
      mark = at position #1 with {
        \arrow{stealth}
      }
    }
  },
  ->-/.default = 0.5
}
\tikzset{
  -<-/.style={
    thick,
    postaction = {decorate},
    decoration = {
      markings,
      mark = at position #1 with {
        \arrow{stealth reversed}
      }
    }
  },
  -<-/.default = 0.5
}

\newcommand\drawFileIcon[9]{
  \def\corner{#1}
  \def\cornerRadius{#2}
  \def\lWidth{#3}
  \def\h{#4}
  \def\w{#5}
  \def\nLines{#6}
  \def\lMargin{#7}
  \def\tMargin{#8}
  \def\startPosition{#9}
  \drawFileIconContinued
}
\newcommand\drawFileIconContinued[5]{
  \coordinate (nw) at ($(\startPosition)+(#1,\h/2)-(\w/2,#2)$);
  \coordinate (ne0) at ($(nw) + (\w, 0)$);
  \coordinate (ne1) at ($(ne0) - (\corner, 0)$);
  \coordinate (ne2) at ($(ne0) - (0, \corner)$);
  \coordinate (se) at ($(ne0) + (0, -\h)$); 
  \filldraw [-, line width = \lWidth, fill=white] (nw) -- (ne1) -- (ne2) [rounded corners=\cornerRadius]--(se) -- (nw|-se) -- cycle;
  \draw [-, line width = \lWidth] (ne1) [rounded corners=\cornerRadius]-- (ne1|-ne2) -- (ne2);
  \node [text width = \w, minimum height = \tMargin, anchor = north west] at (nw) {\tt #4};
  \if\nLines0
    \node [text width = \w, minimum height = \h-\tMargin, anchor = north west] at ($(nw) + (0, -\tMargin)$) {\tt #5};
  \else
    \foreach \k in {1,...,\nLines} {
      \draw [-, line width = \lWidth, line cap=round]
      ($(nw|-se) + (\lMargin,\lMargin) + (0,{(\k-1)/(\nLines-1)*(\h - \lMargin - \tMargin)})$)
      -- ++ ($(\w,0) - 2*(\lMargin,0)$);
    }
  \fi
  \path (nw)+(\w/2,0-\h/2) node [minimum width=\w, minimum height=\h] (#3) {}; 
}

\newcommand{\gear}[7]{
  (0:#2)
  \foreach \i [evaluate=\i as \n using {\i-1)*360/#1}] in {1,...,#1}{
    arc (\n:\n+#4:#2) {[rounded corners=1.5pt] -- (\n+#4+#5:#3)
    arc (\n+#4+#5:\n+360/#1-#5:#3)} --  (\n+360/#1:#2)
  };
  \node[circle, text width=#2*55, align=center, inner sep=0pt] (#6) {#7}
}

\def\checkmark{\tikz\fill[scale=0.4](0,.35) -- (.25,0) -- (1,.7) -- (.25,.15) -- cycle;}

\makeatletter
\newcommand{\op}[3][\@empty]{
  \ifx\@empty#1\relax
    \mathbf{#2}_{#3}
  \else
    \overset{\bullet}{\mathbf{#2}}_{#3}
  \fi
}

\newcommand{\ALPHA}{\large\boldsymbol{\alpha}\normalsize}
\newcommand{\GAMMA}{\large\boldsymbol{\gamma}\normalsize}
\newcommand{\mt}{\mathcal}
\DefineVerbatimEnvironment{snippet}{Verbatim}{
  frame=single,
  framerule=0.1mm,
  rulecolor=\color{gray},
  labelposition=bottomline,
  numbers=left,
  numbersep=2pt,
  numberblanklines=true,
  commandchars=\\\{\},
  framesep=2mm,
  xleftmargin=0mm,
  xrightmargin=0mm,
  label={\bf\snippetLabel},
  codes={\catcode`\$=3\catcode`^=7\catcode`_=8}
}

\hypersetup{
  pdfstartview = {FitH},
  pdffitwindow = true,
  pdfnewwindow = true,
  colorlinks   = true,
  urlcolor     = blue,
  linkcolor    = blue,
  citecolor    = blue,
  pdfauthor    = {Dibyendu Das},
  pdfcreator   = {LaTeX},
  pdfkeywords  = {Interval Analysis, Static Analysis, Abstract Interpretation, Probabilistic Programs},
  pdftitle     = {Probabilistic Interval Analysis of Unreliable Programs},
  pdfsubject   = {Research Paper},
  pdfpagemode  = UseNone
}

\newtheorem{theorem}{Theorem}[section]
\newtheorem{definition}{Definition}[section]

\newtheorem{lemma}[theorem]{Lemma}

\newtheorem{example}[theorem]{Example}

\author[D. Das and S. Dey]
    {Dibyendu Das$^1$ and Soumyajit Dey$^1$\\
     $^1$Department of Computer Science \& Engineering,\\
     Indian Institute of Technology Kharagpur, India}

\correspond{Dibyendu Das, Department of Computer Science \& Engineering, Indian Institute of Technology Kharagpur, Kharagpur, West Bengal 721302, India. e-mail: dibyendu.das.in@gmail.com}

\pubyear{2024}
\pagerange{\pageref{firstpage}--\pageref{lastpage}}

\begin{document}
\label{firstpage}

\makecorrespond


\noindent{\fontsize{22pt}{22pt}\selectfont
Probabilistic Interval Analysis of Unreliable Programs}\\

\noindent{\fontsize{16pt}{16pt}\selectfont
Dibyendu Das$^1$\ \ and\ \ Soumyajit Dey$^1$}\\
\noindent{\fontsize{14pt}{14pt}\selectfont
$^1$Department of Computer Science \& Engineering,}\\
\noindent{\fontsize{14pt}{14pt}\selectfont
Indian Institute of Technology Kharagpur, India}
\vspace{2cm}

\begin{abstract}
Advancement of chip technology will make future computer chips faster. Power consumption of such chips shall also decrease. But this speed gain shall not come free of cost, there is going to be a trade-off between speed and efficiency, i.e accuracy of the computation. In order to achieve this extra speed we will simply have to let our computers make more mistakes in computations. Consequently, systems built with these type of chips will possess an innate unreliability lying within. Programs written for  these systems will also have to incorporate this unreliability. Researchers have already started developing programming frameworks for unreliable architectures as such.

In the present work, we use a restricted version of \emph{C-type} languages to model the programs written for unreliable architectures. We propose a technique for statically analyzing codes written for these kind of architectures. Our technique, which primarily focuses on \emph{Interval/Range Analysis} of this type of programs, uses the well established theory of \emph{abstract interpretation}. While discussing unreliability of hardware, there comes scope of failure of the hardware components implicitly. There are two types of failure models, namely: 
\begin{inparaenum}[\itshape 1\upshape)]
  \item \emph{permanent failure model}, where the hardware stops execution on failure and
  \item \emph{transient failure model}, where on failure, the hardware continues subsequent operations with wrong operand values.
\end{inparaenum}
In this paper, we've only taken transient failure model into consideration. The goal of this analysis is to predict   {\em  the probability with which a program variable assumes values from a given range at a given program point}. 

\end{abstract}


\section{Introduction}
As transistors approach sub-atomic sizes, their functional reliability is increasingly compromised primarily due to particle effects triggered by high temperature footprints over very small area. Computer-chip designers keep figuring out technical workarounds to the miniaturization problem; however current manufacturing techniques have ran out of steam, and Moore’s Law --- the prediction that has yielded huge increases in semiconductor performance for nearly five decades do not hold any more.


Emerging high-performance architectures are anticipated to contain unreliable components that may exhibit soft errors, which silently corrupt the results of computations. While full detection and masking of soft errors is challenging, expensive, and, for some applications, unnecessary, some applications can withstand a certain amount of errors. For example, approximate computing applications such as
\begin{inparaenum}[\itshape i\upshape)]
  \item Multimedia processing,
  \item High definition video decoding,
  \item Image manipulation,
  \item Machine learning,
  \item Big data analytics
\end{inparaenum}
can often naturally tolerate soft errors.

Works have already been done on modeling and manufacturing unreliable hardware components. Probabilistic CMOS or PCMOS technology yields several orders of magnitude improvements in terms of energy over conventional (deterministic) CMOS based implementations~\cite{chakrapani08}. Foundational principles of PCMOS technology are used to build probabilistic switches, which have been used to realize FIR filters as part of the H.264 decoding algorithm~\cite{marpe05}. Similar effort is the work on Stochastic processors that tolerate hardware errors while producing outputs that are, in the worst case, stochastically correct. Analog computers based on a regular array of stochastic computing element  logic have been reported in ~\cite{rascel69}. For capturing such computing inaccuracies under a formal framework, a new programming framework  called RELY has been proposed in ~\cite{carbin13} that enables software developers to specify tolerable error bounds.


We propose a general \emph{abstract interpretation}~\cite{cousot77}~\cite[p.~211]{nielson99} based method for statically analyzing programs that run on unreliable hardware. Abstract interpretation is a well established theory for program analysis and it is already in use for imperative programs that run on reliable hardware. Our contribution in this paper is to extend that theory for unreliable programs i.e programs that run on unreliable architecture. In this paper, we present an \emph{Probabilistic} Interval Analysis (or Range Analysis) for programs that are supposed to run on  unreliable hardware. Our approach takes into count, the unreliability that occurs because of Arithmetic, Logical and Memory operations generally found in all high level programming languages. It performs interval or range analysis~\cite{harrison77} for programs like these. The objective is to calculate the reliability of each program variable at each program point as well as the correctness probability of the overall program.  In summary, our contributions can be itemized as follows. 

\begin{itemize}
  \item We propose a technique for \emph{Probabilistic Range/Interval Analysis} of C programs. Our idea is centered around creation of a  \emph{probabilistic interval abstract domain} and  defining \emph{widening} operations for such domains.
 \item We have implemented our theory in the form of a prototype tool that takes a \emph{C-like} program and a hardware reliability specification from the user. It statically analyses the program using our theory of probabilistic interval analysis in view of the reliability specification and finally gives an output of operating intervals associated with corresponding probabilities for each variable at each program point.
  \end{itemize}


\section{Interval Analysis of Unreliable Programs}

\begin{figure}
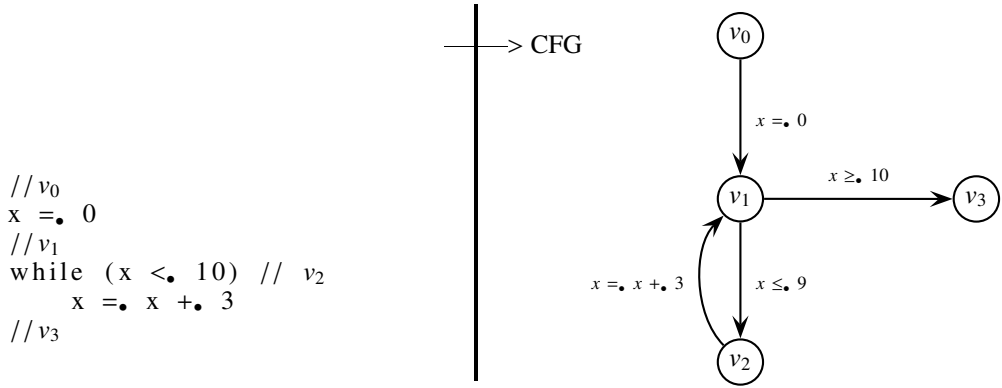

  \begin{minipage}[b]{0.40\linewidth}
    \begin{lstlisting}[mathescape]
    //$v_0$
    x $=_\bullet$ 0
    //$v_1$
    while (x $<_\bullet$ 10) // $v_2$
        x $=_\bullet$ x $+_\bullet$ 3
    //$v_3$
    \end{lstlisting}
  \end{minipage}
  \rule[42mm]{4mm}{.1pt}\rule[-2.5mm]{1pt}{50mm}\rule[42mm]{4mm}{.1pt}\raisebox{41.2mm}{$>$ CFG}
  \hspace{0.1\linewidth}
  \begin{minipage}[b]{0.40\linewidth}
    $
    \psmatrix[colsep=2.5cm,rowsep=1.5cm,mnode=circle]
      v_0       \\
      v_1 & v_3 \\
      v_2
      \psset{arrowscale=2}
      \everypsbox{\scriptstyle}
      \ncline{->}{1,1}{2,1}>{x\ =_\bullet\ 0}
      \ncline{->}{2,1}{2,2}^{x\ \geq_\bullet\ 10}
      \ncline{->}{2,1}{3,1}>{x\ \leq_\bullet\ 9}
      \ncarc[arcangle=-45]{<-}{2,1}{3,1}<{x\ =_\bullet\ x\ +_\bullet\ 3}
    \endpsmatrix
    $
  \end{minipage}
  \caption{Example program}
  \label{example-program-figure} 
\end{figure}

We'll start by defining what unreliable programs are in this context. To do so, we shall take an example of a sample C-like program with appropriate modifications wherever necessary. Fig.~\ref{example-program-figure} shows an example unreliable program snippet and its corresponding \emph{Control Flow Graph}. 
This program is subject to execution on an unreliable system in which every operation (ALU or Memory) is unreliable. In this example program each operation is probabilistic i.e each operation produces correct result with some probability. Unreliable operators are denoted by the corresponding operator followed by a \emph{subscript bullet} ($\bullet$). Each operator has a success probability associated with it i.e the operator succeeds and produces correct result with a predefined probability. These probability values are set by the hardware manufacturer of the chip. Upon failure the probabilistic operators take random values from their corresponding domains. For arithmetic and memory operations this domain is $\{i\in\mathbb{Z}\ |\ \mathtt{MININT}\leq i\leq\mathtt{MAXINT}\}$\footnote{We've assumed that each integer variable in the program takes values between two limiting boundary values $\mathtt{MININT}$ and $\mathtt{MAXINT}$}, for boolean operators the domain of operation is $\{true, false\}$. Success and corresponding failure probabilities of some of the operators are given in Table~\ref{hw-spec}.
\begin{table}
  \caption{Hardware Probability Specification}
  \begin{tabular}{@{\hspace{5mm}}c@{\hspace{1cm}}c@{\hspace{1cm}}c@{\hspace{1cm}}l@{}}

   \multirow{2}{*}{Operator} & Probability of successful execution & \multirow{2}{*}{Failure probability} & \multirow{2}{*}{Domain of operation} \\
   & (set by hardware manufacturer) & & \\

   $+_\bullet$ & $Pr(+_\bullet)$ & $1-Pr(+_\bullet)$ & \multirow{5}{*}{$\big\{a\in\mathbb{Z}\ |\ \mathtt{MININT}\leq a\leq\mathtt{MAXINT}\big\}$} \\
   $-_\bullet$ & $Pr(-_\bullet)$ & $1-Pr(-_\bullet)$ & \\
   $\times_\bullet$ & $Pr(\times_\bullet)$ & $1-Pr(\times_\bullet)$ & \\
   $\div_\bullet$ & $Pr(\div_\bullet)$ & $1-Pr(\div_\bullet)$ & \\
   $=_\bullet$ & $Pr(Wr)$ & $1-Pr(Wr)$ & \\

   $>_\bullet$ & $Pr(>_\bullet)$ & $1-Pr(>_\bullet)$ & \multirow{2}{*}{$\big\{true,\ false\big\}$} \\
   $\geq_\bullet$ & $Pr(\geq_\bullet)$ & $1-Pr(\geq_\bullet)$ & \\

   $\vdots$ & $\vdots$ & $\vdots$ &
  \end{tabular}
  \label{hw-spec}
\end{table}

\par We've modeled unreliable programs with the syntactic rules of the following BNF. It is a context-free grammar to capture  C-type programs with an exception of the newly added unreliable operators. Unreliable operators are denoted by the corresponding operator followed by a \emph{subscript bullet} ($\bullet$). Unreliable operators are introduced in the production rules for the non-terminal \synt{expr} used to represent generic expressions in the BNF.

\setlength{\grammarparsep}{6pt} 
\setlength{\grammarindent}{10em} 
\begin{grammar}



















<expr> ::= `IDENT' \lit{$=_\bullet$} <expr>
\alt <expr> \lit{$\|_\bullet$} <expr>
\alt <expr> \lit{$\&\&_\bullet$} <expr>
\alt <expr> \lit{$==_\bullet$} <expr>\quad|\quad<expr> \lit{$\neq_\bullet$} <expr>
\alt <expr> \lit{$\leqslant_\bullet$} <expr>\quad|\quad<expr> \lit{$\textless_\bullet$} <expr>\quad|\quad<expr> \lit{$\geqslant_\bullet$} <expr>\quad|\quad<expr> \lit{$\textgreater_\bullet$} <expr>
\alt <expr> \lit{$+_\bullet$} <expr>\quad|\quad<expr> \lit{$-_\bullet$} <expr>
\alt <expr> \lit{$\times_\bullet$} <expr>\quad|\quad<expr> \lit{$\div_\bullet$} <expr>\quad|\quad<expr> \lit{$\%_\bullet$} <expr>
\alt \lit{$!_\bullet$} <expr>\quad|\quad\lit{$-_\bullet$} <expr>\quad|\quad\lit{$+_\bullet$} <expr>
\alt `(' <expr> `)'


\end{grammar}
We take a sample program statement `$\sigma : x\ =_\bullet\ x\ +_\bullet\ 3$' as an example and determine the overall probability of getting the correct result after the execution of the statement. The program statement $\sigma$ involves 3 probabilistic operations namely Reading (\textbf{Rd}) of the variable $x$, Addition (\textbf{+}) \& Writing back (\textbf{Wr}) to the variable $x$. So, the probability that $\sigma$ executes successfully is $\quad Pr(Rd)\cdot Pr(+_\bullet)\cdot Pr(Wr)$ with $Pr$ denoting the success probability of the operation concerned. 

We are interested in computing the probability with which a program variable $x$ holds the correct value after the execution of $\sigma$ for a fault model defined as follows. If any operation, e.g. addition ($+_\bullet$ say) succeeds, it will produce the  correct result with  probability $Pr(+_\bullet)$. But if it fails (with probability $1-Pr(+_\bullet)$), it can randomly produce  any result from its  domain, given by $[\mathtt{MININT}, \mathtt{MAXINT}] = \left\{ i \in \mathbb{N} \ \middle|\ \mathtt{MININT} \leq i \leq  \mathtt{MAXINT} \right\}$. Of all these values in its domain, there exists  one value which itself would have been the correct value of this operation, had it succeeded. Therefore, even if the operation fails, it still produces correct result with a minuscule probability of $\displaystyle\frac{1-Pr(+_\bullet)}{\mathtt{MAXINT}-\mathtt{MININT}+1}$\footnote{There are $\mathtt{MAXINT}-\mathtt{MININT}+1$ number of integers in the domain.}. Therefore, the unreliable arithmetic operation, addition, produces correct result with probability $\displaystyle\left(Pr(+_\bullet) + \frac{1-Pr(+_\bullet)}{\mathtt{MAXINT}-\mathtt{MININT}+1}\right)$. Similar argument is applicable for memory operations (Rd \& Wr) too. But in case of logical operations, the domain becomes $\{true, false\}$. Therefore, an unreliable logical operation, say $\geq_\bullet$, produces correct result with probability $\displaystyle\left(Pr(\geq_\bullet) + \frac{1-Pr(\geq_\bullet)}{2}\right)$\footnote{If a logical operator fails, it is equally likely to choose any value from its domain, $\{true, false\}$. $\therefore$ upon failure, a logical operator, say $\geq_\bullet$, still gives correct result with probability $\displaystyle\frac{1-Pr(\geq_\bullet)}{2}$}.

Let $Pr_{post, \sigma}(x)$ / $Pr_{pre, \sigma}(x)$ denote the probability of $x$ holding the correct value after/before the execution of $\sigma$. Assuming that all operations involved in $\sigma$ are independent of each other, we have,
\begin{multline}
  Pr_{post, \sigma}(x) = Pr_{pre, \sigma}(x)\cdot\left(Pr(Rd) + \frac{1-Pr(Rd)}{\mathtt{MAXINT}-\mathtt{MININT}+1}\right)\cdot\left(Pr(+_\bullet) + \frac{1-Pr(+_\bullet)}{\mathtt{MAXINT}-\mathtt{MININT}+1}\right)\cdot \\ \left(Pr(Wr) + \frac{1-Pr(Wr)}{\mathtt{MAXINT}-\mathtt{MININT}+1}\right) \nonumber
\end{multline}

In the following sections we follow the general principles of Abstract interpretation~\cite[p.~211]{nielson99}
in order to construct a \emph{Probabilistic Concrete Domain}, a \emph{Probabilistic Interval  Abstract Domain}, establish a \emph{Galois Connection} between the domains, and finally define and  approximate a suitable \emph{strongest post-condition function} in the line of ~\cite[sec.~6, pp.~242--243]{cousot77}.

\section{The Probabilistic Concrete Domain, $(\mathcal{L},\sqsubseteq_L)$}
Our idea of probabilistic concrete domain is essentially an extension of standard concrete domain of programs. Given a program  $\sigma$, its concrete domain implies the collection of possible program states at each program point of $\sigma$. In the same vein, a probabilistic concrete domain implies the set of possible program states at each program point of $\sigma$ along with the probabilities of assuming variable values from the collection of possible program states. We represent a probabilistic program state as a tuple of set of values and its associated probability. This is the minimum probability with which a program variable will take its values from the set. For program with a single integer variable, we define the concrete domain, $\mt{L}$, as follows.
\begin{definition} Let $\langle S, p\rangle$ denote that at a particular program point, a program variable takes its value  from set $S$ with probability $p$. $\mt{L}=\left\{\left(S,\ p \right)\ \middle|\ S \in 2^{[\mathtt{MININT},\ \mathtt{MAXINT}]} \bigwedge 0\leq p\leq1\right\}$.  In general, for programs with $n$ integer variables, the concrete domain will be $\mt{L}^n$.
\end{definition}
 We define a partial order $\sqsubseteq_L$ on $\mt{L}$ as follows :
\begin{definition}\label{concrete_order}
$\forall \langle S_1,p_1\rangle, \langle S_2,p_2\rangle\in \mt{L}\qquad\langle S_1,p_1\rangle\sqsubseteq_L \langle S_2,p_2\rangle\ \iff \  S_1\subseteq S_2\ \bigwedge\ p_1\geq p_2$
\end{definition}
The intuition behind the definition can be explained as follows. The element $\langle S_2,p_2\rangle$ which is higher in the order is an abstraction of $\langle S_1,p_1\rangle$ such that the element $\langle S_1,p_1\rangle$ provides more precise information about program state since a variables value is predicted to be in a smaller range with higher probability.

\begin{lemma}
$(\mt{L},\ \sqsubseteq_L)$ is a \textbf{poset}.
\begin{proof}
We prove that $\sqsubseteq_L$ is a partial order relation by showing reflexivity, transitivity and anti-symmetricity for $\sqsubseteq_L$. 
\begin{description}
\item[Reflexivity :] Reflexivity follows trivially from def.~\ref{concrete_order} since $\forall \langle S,p\rangle\in \mt{L}, \langle S,p\rangle\sqsubseteq_L\langle S,p\rangle$.
\item[Transitivity :] Given $\langle S_1,p_1\rangle\sqsubseteq_L\langle S_2,p_2\rangle$ and $\langle S_2,p_2\rangle\sqsubseteq_L\langle S_3,p_3\rangle$, we have,
\begin{align*}
S_1\subseteq S_2 &\ \bigwedge\ S_2\subseteq S_3\quad\implies\quad S_1\subseteq S_3\\
p_1\geq p_2 &\ \bigwedge\ p_2\geq p_3\quad\implies\quad p_1\geq p_3
\end{align*}
Hence, $\langle S_1,p_1\rangle\sqsubseteq_L\langle S_3,p_3\rangle$. 
\item[Anti-Symmetricity :] Given  $\langle S_1,p_1\rangle\sqsubseteq_L\langle S_2,p_2\rangle$ and $\langle S_2,p_2\rangle\sqsubseteq_L\langle S_1,p_1\rangle$, we have, 
\begin{align*}
    S_1\subseteq S_2 &\ \bigwedge\ S_2\subseteq S_1\quad\implies\quad S_1=S_2\\
    p_1\geq p_2 &\ \bigwedge\ p_2\geq p_1\quad\implies\quad p_1=p_2
\end{align*}
Hence, $\langle S_1,p_1\rangle=\langle S_2,p_2\rangle$.
\end{description}
\end{proof}
\end{lemma}
 

\begin{lemma}
$(\mt{L},\ \sqsubseteq_L)$ is a \textbf{complete lattice}.
\begin{proof}
For proving $(\mt{L},\ \sqsubseteq_L)$ to be a \textbf{\emph{lattice}}, we need to show that for any two elements in $\mt{L}$, there exist a unique infimum (greatest lower bound) and a unique supremum (least upper bound). 
The least upper bound $(l.u.b)$ operator, $\displaystyle\bigsqcup_L$ and the greatest lower bound $(g.l.b)$ operator,  $\displaystyle\bigsqcap^L$ for any two elements $\langle S_1,p_1\rangle, \langle S_2,p_2\rangle \in \mt{L}$ can be defined as follows.

\begin{align}
\langle S_1,p_1\rangle\ \bigsqcup_L\ \langle S_2,p_2\rangle\quad=\quad&\left\langle S_1\bigcup S_2, \textbf{min}(p_1,p_2)\right\rangle \in \mt{L} &\label{eq:lub_L}\\
\langle S_1,p_1\rangle\ \bigsqcap^L\ \langle S_2,p_2\rangle\quad=\quad&\left\langle S_1\bigcap S_2, \textbf{max}(p_1,p_2)\right\rangle \in \mt{L}  &\label{eq:glb_L}
\end{align}
Soundness of these definitions can be established as follows. From eq.~\ref{eq:lub_L}, we get the least upper bound, $\langle S_u,p_u\rangle$ of $\langle S_1,p_1\rangle$ and $\langle S_2,p_2\rangle$ as $\displaystyle\langle S_u,p_u\rangle = \langle S_1,p_1\rangle \bigsqcup_L \langle S_1,p_1\rangle = \left\langle S_1\bigcup S_2,\textbf{min}(p_1,p_2)\right\rangle$. Now, let $\langle S_c,p_c\rangle$ be any upper bound of $\langle S_1,p_1\rangle$ and $\langle S_2,p_2\rangle$. Hence, $\langle S_1,p_1\rangle\sqsubseteq_L\langle S_c,p_c\rangle$  and $\langle S_2,p_2\rangle\sqsubseteq_L \langle S_c,p_c\rangle$. Note that, 
\begin{align*}
&S_1\subseteq S_c\ \bigwedge\ S_2\subseteq S_c\quad\implies\quad S_1\bigcup S_2\subseteq S_c&\\
&p_1\geq p_c\ \bigwedge\ p_2\geq p_c\quad\implies\quad \textbf{min}(p_1,p_2)\geq p_c&\\
\therefore\quad&\left\langle S_1\bigcup S_2, \textbf{min}(p_1,p_2)\right\rangle\quad\sqsubseteq_L\quad\langle S_c,p_c\rangle&\hfill\text{[from def.~\ref{concrete_order}]}\\
\Rightarrow\quad&\langle S_u,p_u\rangle\ \sqsubseteq_L\ \langle S_c,p_c\rangle&
\end{align*}
Thus $\langle S_u,p_u\rangle$ is the least upper bound of $\langle S_1,p_1\rangle$ and $\langle S_2,p_2\rangle$. Similarly, a proof of soundness for the greatest lower bound operator $\displaystyle\bigsqcap^L$ can also be obtained. The greatest and least element of $(\mt{L},\ \sqsubseteq_L)$ are $\top_L$ and $\bot_L$ respectively and they are defined as, 
$\top_L=\langle\{i\in\mathbb{Z}\ |\ \mathtt{MININT}\leq i\leq\mathtt{MAXINT}\},\ 0\rangle$ and $\bot_L=\langle\{\ \}, 1\rangle$. Hence, $\displaystyle\left\langle \mt{L}, \sqsubseteq_L, \bigsqcup_L, \bigsqcap^L \right\rangle$ is a lattice with greatest and least elements $\top_L$ and  $\bot_L$ respectively. The operations $\displaystyle\bigsqcup_L \text{and} \bigsqcap^L$ can be extended over any $\mt{S} = \{\langle S_1, p_1\rangle, \cdots , \langle S_n, p_n\rangle\}\subseteq \mt{L}$ such that $\mt{S}$ has a unique $l.u.b$ and $g.l.b$ in $\mt{L}$ given by $\displaystyle\bigsqcup_L \mt{S} = \left\langle \bigcup_{i=1}^{n} S_i,\ \textbf{min}(p_1,\cdots,p_n) \right\rangle$ and $\displaystyle\bigsqcap_L \mt{S} = \left\langle \bigcap_{i=1}^{n} S_i,\ \textbf{max}(p_1,\cdots,p_n) \right\rangle$ respectively. Hence $\mt{L}$ is a complete lattice.
\end{proof}	
\end{lemma}

\subsection{Hasse Diagram of Lattice $(\mt{L},\ \sqsubseteq_L)$}

\begin{figure}
  \centering
  \resizebox{\linewidth}{!}{
    \begin{tikzpicture}[every node/.style={draw, circle, inner sep=0pt, outer sep=0pt, font=\small}, scale=0.45]
	  \input{figures/concrete_frontal.tex}
    \end{tikzpicture}
  }
  \caption{Frontal View of $(\mt{L},\sqsubseteq_L)$}
  \label{concrete-frontal-figure}
\end{figure}

The lattice $(\mt{L},\sqsubseteq_L)$ is a lattice over continuous intervals of real numbers, so it has chains of infinite length. Its Hasse diagram is best viewed in 3 dimensions. For that reason, we've shown several different projectional views of the same diagram. In all the projectional views, \emph{directed solid lines} represent $g.l.b\to l.u.b$ relationships i.e of the two elements connected by such an edge, the element at the arrow head is the least upper bound and the element at the arrow base is the greatest lower bound. There doesn't exist any other element between them. Whereas \emph{directed dashed lines} represent the existence of infinite number of elements between the connecting pair. Along any dashed edge, the set of values remains the same but probability decreases monotonically.

In fig.~\ref{concrete-frontal-figure}, fig.~\ref{concrete-lateral-figure} and fig.~\ref{concrete-rear-figure}, we show respectively the frontal view, a skewed lateral view and the rear view of the lattice. The legends used in the diagrams are explained as follows:

\begin{description}
  \item[In fig.~\ref{concrete-frontal-figure} (Frontal View):] \hfill
    \begin{itemize}
      \item $^{^{A}}\mathbb{Z}_{_{B}}=\left\{i\in\mathbb{Z}\ |\ A\leq i\leq B\right\}$
    \end{itemize}
  \item[In fig.~\ref{concrete-frontal-figure}, fig.~\ref{concrete-lateral-figure} and fig.~\ref{concrete-rear-figure}:] \hfill
    \begin{itemize}
      \item $m=\mathtt{MININT}$
      \item $M=\mathtt{MAXINT}$
      \item $A_{Ri}\equiv A_{(M-m)i}\quad\text{and}\quad\hat{A}_{Ri}\equiv\hat{A}_{(M-m)i}\qquad\forall i$
      \item For every pair of indices $i, j$ (and $\top$) the elements denoted by $A_{ij}$ (and $A_\top$) and $\hat{A}_{ij}$ (and $\hat{A}_\top$) have the same set of values, but probability in element $A_{ij}$ (and $A_\top$) is $0$ , where as it is 1 for $\hat{A}_{ij}$ (and $\hat{A}_\top$).\\
      E.g: $A_{22}\equiv\left<^{^{m+1}}\mathbb{Z}_{_{M-1}},\ 0\right>$ but $\hat{A}_{22}\equiv\left<^{^{m+1}}\mathbb{Z}_{_{M-1}},\ 1\right>$
    \end{itemize}
\end{description}

\begin{figure}
  \centering
  \resizebox{\linewidth}{!}{
    \begin{tikzpicture}[x={(6mm,-3.5mm)}, y={(0mm,15mm)}, z={(14mm,0mm)}, every node/.style={inner sep=2pt}, scale=0.8]
      \input{figures/concrete_lateral.tex}
    \end{tikzpicture}
  }
  \caption{Lateral View (\textit{skewed}) of $(\mt{L},\sqsubseteq_L)$}
  \label{concrete-lateral-figure}
\end{figure}

\begin{figure}
  \centering
  \resizebox{\linewidth}{!}{
    \begin{tikzpicture}
      \input{figures/concrete_rear.tex}
    \end{tikzpicture}
  }
  \caption{Rear View of $(\mt{L},\sqsubseteq_L)$}
  \label{concrete-rear-figure}
\end{figure}

Now that we have defined our concrete domain $\mt{L}^n$ of program variables, following the line of abstract interpretation, the next thing we require is to define a function on $\mt{L}^n$ that modifies the program states in the same way it would have been modified had the program been actually executed. In the next section we define one such function over $\mt{L}^n$ called the \emph{strongest post condition function}.

\subsection{Strongest Post-Condition Function, $sp : \mt{L}^n\to\mt{L}^n$}

Before defining the strongest post condition function, here is a brief overview of what pre- and post-conditions are. Pre- and post-conditions are conditions (assertions) on program variables that hold respectively before and after the execution of the program (or a particular segment of the program). Usually pre- and post-conditions are represented as set of predicates defined over the program variables. For any program $\sigma$, its pre- and post-conditions are expressed in the form of \emph{Hoare Logic}, i.e $\{P\}\sigma\{Q\}$, where, $P$ is the set of pre-conditions and $Q$ is the set of post-conditions.

Given a set of pre-conditions $P$ and a program $\sigma$, a \emph{post-condition function} is a function, that generates the set of post-conditions $Q$ i.e the conditions that will hold after the program $\sigma$ actually executes. For any predicate $p\in P$, defined over program variables, we can alternatively think of $p$ as a set of possible program states which satisfy the relation $p$. Let's explain it with an example as follows:

\begin{example}
Given an integer variable $x$, consider $p:\langle\{5\},p_1\rangle\footnote{$p_1$ is just a place holder here}$ to be a program state which represents the fact that `$x==5$' (i.e $p\equiv[x=5]$) and $\mathfrak{p}:[x>10]$ to be a predicate such that $p\notin \mathfrak{p}$. Similarly, consider a program with three integer variables $x,y$ and $z$ such that $p:\langle(\{5\},p_x),(\{10\},p_y),(\{10\},p_z)\rangle\footnote{$p_x,p_y,p_z$ are all place holders}\equiv[x=5,\ y=10,\ z=10]$ and $p':\langle(\{15\},p_x),(\{0\},p_y),(\{1\},p_z)\rangle\equiv[x=15,\ y=0,\ z=1]$ are two program states. Given a predicate $\mathfrak{p}:[x+y\geq z]$, we may observe that both $p,p'\in \mathfrak{p}$.
\end{example}

Thus, henceforth we'll think of the set of pre- and post-conditions as sets of equivalent program states.

Let, $p$ denotes a pre-condition predicate which is true before the execution of some program $\sigma$. Strongest post condition, $sp$, is a function which takes as arguments a program statement $\sigma$ (or a sequence of statements) and a predicate $p$ and returns the strongest predicate that holds after executing $\sigma$ from any program state which satisfies $p$. It is strongest in the sense that for any other such predicate $q$ which holds for any states resulting from execution of $\sigma$ given $p$, we shall always have $sp(p,\sigma)\Rightarrow q$ or in a set theoretic notation $sp(p,\sigma)\subseteq q$.

Now that we have in our hand the basic definitions required for giving a more general form of the \emph{strongest post condition function}, let's go ahead and define the same as follows:

We assume that there are $n$ variables in the program, namely, $x_1,x_2\cdots x_n$. At any program point, a predicate with $n$ variables will be of the form $P:\langle(S_1,p_1),(S_2,p_2),\cdots,(S_n,p_n)\rangle$, s.t at that program point, each $x_i$ will assume values from the set $S_i$ with probability $p_i$. Before giving the definitions of $sp$ for arithmetic and logical expressions, let's define some utility functions that we'll be using the the actual definitions and also afterwards. For a tuple of $n$ integers, $(v_1, v_2, \cdots, v_n)$ and for any program variable $x_i$,

$\mathcal{V}\Big((v_1,v_2,\cdots,v_n),x_i\Big)=
\begin{cases}
    v_i      & \quad \text{if } x_i \text{ is a variable}\\
    x_i=C    & \quad \text{if } x_i \text{ is a constant, say } C\\
\end{cases}$

\noindent and for any program state $P:\langle(S_1,p_1),(S_2,p_2),\cdots,(S_n,p_n)\rangle\in\mt{L}^n$,

$\mathcal{P}(P,x_i)=
\begin{cases}
p_i    & \quad \text{if } x_i \text{ is a variable}\\
1      & \quad \text{if } x_i \text{ is a constant}\\
\end{cases}$

\noindent We'll also be using some symbolic notations in our definition of $sp$, which are as follows:
\begin{flalign}
  &\forall i\in\mathbb{N},\quad\op[prob]{A}{i} \text{ is a probabilistic arithmetic operation}\quad\text{and}\quad\op{A}{i} \text{ is a deterministic arithmetic operation.}&&&\nonumber\\
  &\text{i.e}\quad\op[prob]{A}{i}\in\{+_{\bullet},-_{\bullet},\times_{\bullet},\div_{\bullet},\%_{\bullet},\cdots\}\quad\text{and}\quad\op{A}{i}\in\{+,-,\times,\div,\%,\cdots\}&&&\nonumber\\
  &\op[prob]{C}{} \text{ is a probabilistic comparison operation}\quad\text{and}\quad\op{C}{} \text{ is a deterministic comparison operation.}&&&\nonumber\\
  &\text{i.e}\quad\op[prob]{C}{}\in\{<_{\bullet},>_{\bullet},\geq_{\bullet},\leq_{\bullet},==_{\bullet},\neq_{\bullet}\}\quad\text{and}\quad\op{C}{}\in\{>,<,\geq,\leq,==,\neq\}&&&\nonumber
\end{flalign}

We define the \emph{strongest post-condition function} on $\mt{L}^n$ w.r.t \textbf{arithmetic expressions} (with assignment) as follows:

\begin{definition}\label{sp_arith}
  Let $P$ be a precondition on program states and program statement $\sigma$ be of the form\\
$\left(x_{i_f} =_\bullet x_{i_1}\ \ \op[prob]{A}{2}\ \ x_{i_2}\ \ \op[prob]{A}{3}\ \ \cdots\ \ \op[prob]{A}{k}\ \ x_{i_k}\right)$, where $\forall j,\ 1\leq i_j\leq n$. After executing $\sigma$, the strongest post-condition on program states with be $P^*$ such that,
  \begin{flalign}
    P^*=&\quad sp\Big(P,\ x_{i_f} =_\bullet x_{i_1}\ \ \op[prob]{A}{2}\ \ x_{i_2}\ \ \op[prob]{A}{3}\ \ \cdots\ \ \op[prob]{A}{k}\ \ x_{i_k}\Big)&&&\nonumber\\
    =&\quad sp\Big(\langle(S_1,p_1),(S_2,p_2),\cdots,(S_{i_f},p_{i_f}),\cdots,(S_n,p_n)\rangle,\ x_{i_f} =_\bullet x_{i_1}\ \ \op[prob]{A}{2}\ \ x_{i_2}\ \ \op[prob]{A}{3}\ \ \cdots\ \ \op[prob]{A}{k}\ \ x_{i_k}\Big)&&&\nonumber\\
    =&\quad \langle(S_1,p_1),(S_2,p_2),\cdots,(S^*_{i_f},p^*_{i_f}),\cdots,(S_n,p_n)\rangle&&&\nonumber\\
    &\text{where,}\qquad S^*_{i_f}=\left\{\mathcal{V}(t,x_{i_1})\ \ \op{A}{2}\ \ \mathcal{V}(t,x_{i_2})\ \ \op{A}{3}\ \ \cdots\ \ \op{A}{k}\ \ \mathcal{V}(t,x_{i_k})\quad\middle|\quad\forall t\in S_1\times S_2\times\cdots\times S_n\right\}\qquad\text{and}&&&\nonumber\\
    &\qquad\qquad\ \ p^*_{i_f}=\left(Pr(Wr)+\frac{1-Pr(Wr)}{\mathtt{MAXINT}-\mathtt{MININT}+1}\right)\cdot\left(Pr(Rd) + \frac{1-Pr(Rd)}{\mathtt{MAXINT}-\mathtt{MININT}+1}\right)^{|vars|}\cdot&&&\nonumber\\
    &\hspace{6cm}\prod_{\forall\ x_i\ \in\ vars} \mathcal{P}(P,x_i)\ \cdot\ \prod_{j=2}^k\left(Pr\left(\op[prob]{A}{j}\right)+\frac{1-Pr\left(\op[prob]{A}{j}\right)}{\mathtt{MAXINT}-\mathtt{MININT}+1}\right)&&&\nonumber\\
    &\qquad\qquad\text{with,}\qquad vars=\left\{x_{i_j}\ :\ \forall j\in \{1,2,\cdots,k\}\text{ and } x_{i_j}\text{ is a variable}\right\}&&&\nonumber
  \end{flalign}
\end{definition}

Now we define the \textit{strongest post-condition function} on $\mt{L}^n$ w.r.t \textbf{logical expressions} as follows:

\begin{definition}\label{sp_bool}
  Let $P$ be a precondition on program states and program statement $\sigma$ be of the form\\
$\left(x_{i_1}\ \ \op[prob]{A}{i_2}\ \ x_{i_2}\ \ \op[prob]{A}{i_3}\ \ \cdots\ \ \op[prob]{A}{i_k}\ \ x_{i_k}\right)\ \op[prob]{C}{}\ \left(x_{j_1}\ \ \op[prob]{A}{j_2}\ \ x_{j_2}\ \ \op[prob]{A}{j_3}\ \ \cdots\ \ \op[prob]{A}{j_m}\ \ x_{j_m}\right)$, where $\forall p,q\ \ 1\leq i_p,j_q\leq n$. After executing $\sigma$, the strongest post-condition on program states with be $P^*$ such that,
  \begin{flalign}
    P^* = &\quad sp\left(P,\ \left(x_{i_1}\ \ \op[prob]{A}{i_2}\ \ x_{i_2}\ \ \op[prob]{A}{i_3}\ \ \cdots\ \ \op[prob]{A}{i_k}\ \ x_{i_k}\right)\ \op[prob]{C}{}\ \left(x_{j_1}\ \ \op[prob]{A}{j_2}\ \ x_{j_2}\ \ \op[prob]{A}{j_3}\ \ \cdots\ \ \op[prob]{A}{j_m}\ \ x_{j_m}\right)\right)&&&\nonumber\\
    =&\quad sp\left(\langle(S_1,p_1),(S_2,p_2),\cdots,(S_n,p_n)\rangle,\ \left(x_{i_1}\ \ \op[prob]{A}{i_2}\ \ x_{i_2}\ \ \op[prob]{A}{i_3}\ \ \cdots\ \ \op[prob]{A}{i_k}\ \ x_{i_k}\right)\ \op[prob]{C}{}\ \left(x_{j_1}\ \ \op[prob]{A}{j_2}\ \ x_{j_2}\ \ \op[prob]{A}{j_3}\ \ \cdots\ \ \op[prob]{A}{j_m}\ \ x_{j_m}\right)\right)&&&\nonumber\\
    =&\quad \langle(S^*_1,p^*_1),(S^*_2,p^*_2),\cdots,(S^*_n,p^*_n)\rangle&&&\nonumber\\
    &\text{where,}\qquad\forall i\in\{1,2,\cdots,n\},\qquad S^*_i=\left\{t_i\ :\ \forall(t_1,t_2,\cdots,t_n)\in Q\right\}\qquad\text{and}&&&\nonumber\\
    &\qquad\qquad\ \ p^*_i=p_i\cdot\left(Pr(Rd) + \frac{1-Pr(Rd)}{\mathtt{MAXINT}-\mathtt{MININT}+1}\right)^{|vars|}\cdot\left(Pr\left(\op[prob]{C}{}\right)+\frac{1-Pr\left(\op[prob]{C}{}\right)}{2}\right)\ \cdot&&&\nonumber\\
    &\hspace{3cm}\prod_{j=2}^k\left(Pr\left(\op[prob]{A}{i_j}\right)+\frac{1-Pr\left(\op[prob]{A}{i_j}\right)}{\mathtt{MAXINT}-\mathtt{MININT}+1}\right)\cdot\prod_{i=2}^m\left(Pr\left(\op[prob]{A}{j_i}\right)+\frac{1-Pr\left(\op[prob]{A}{j_i}\right)}{\mathtt{MAXINT}-\mathtt{MININT}+1}\right)&&&\nonumber\\
    &\qquad\qquad\text{with,}\qquad vars=\left\{x\in\{x_{i_1},x_{i_2}\cdots x_{i_k}\}\cup\{x_{j_1},x_{j_2}\cdots x_{j_m}\}\ :\ x\text{ is a variable}\right\}\quad\text{and}&&&\nonumber\\
    &\qquad\qquad Q=\left\{t\in S_1\times S_2\times\cdots\times S_n\ \middle|\ \left(\mathcal{V}(t,x_{i_1})\ \ \op{A}{i_2}\ \ \mathcal{V}(t,x_{i_2})\ \ \op{A}{i_3}\ \ \cdots\ \ \op{A}{i_k}\ \ \mathcal{V}(t,x_{i_k})\right)\ \op{C}{}\right.&&&\nonumber\\
    &\hspace{9cm}\left.\left(\mathcal{V}(t,x_{j_1})\ \ \op{A}{j_2}\ \ \mathcal{V}(t,x_{j_2})\ \ \op{A}{j_3}\ \ \cdots\ \ \op{A}{j_m}\ \ \mathcal{V}(t,x_{j_m})\right)\right\}&&&\nonumber
  \end{flalign}
\end{definition}

With this general definition of $sp$ over $\mt{L}^n$, we'll apply it on our example program from fig.~\ref{example-program-figure} and show how the program states are being modified by $sp$ and eventually reaching a fix point of program states.

\subsubsection{Fix point computation of concrete program states using $sp$}

Since our example program from fig.~\ref{example-program-figure} contains only a single integer variable $x$, we simplify the generalized definitions of $sp$ from the previous section i.e def.~\ref{sp_arith} and def.~\ref{sp_bool} in accordance with the program statements used in the example program as follows:
\begin{flalign}
  &sp\Big(\langle S,p\rangle,\ x =_\bullet 0\Big)&=&\left\langle\{0\},\ Pr(Wr)+\frac{1-Pr(Wr)}{\mathtt{MAXINT}-\mathtt{MININT}+1}\right\rangle&\label{eq:sp1}\\
  &sp\Big(\langle S,p\rangle,\ x =_\bullet x +_\bullet 3\Big)&=&\left\langle\left\{v + 3\ :\ \forall v\in S\right\},\ p\cdot\left(Pr(Rd) + \frac{1-Pr(Rd)}{\mathtt{MAXINT}-\mathtt{MININT}+1}\right)\cdot\right.\nonumber&\\
  &&&\left.\left(Pr(+_\bullet) + \frac{1-Pr(+_\bullet)}{\mathtt{MAXINT}-\mathtt{MININT}+1}\right)\cdot\left(Pr(Wr) + \frac{1-Pr(Wr)}{\mathtt{MAXINT}-\mathtt{MININT}+1}\right)\right\rangle&\label{eq:sp2}\\
  &sp\Big(\langle S,p\rangle,\ x \leq_\bullet 9\Big)&=&\left\langle\left\{v\in S\ :\ v\leq 9\right\},\ p\cdot\left(Pr(\leq_\bullet) + \frac{1-Pr(\leq_\bullet)}{2}\right)\cdot\right.&\nonumber\\
  &&&\hspace{5cm}\left.\left(Pr(Rd) + \frac{1-Pr(Rd)}{\mathtt{MAXINT}-\mathtt{MININT}+1}\right)\right\rangle&\label{eq:sp3}\\
  &sp\Big(\langle S,p\rangle,\ x \geq_\bullet 10\Big)&=&\left\langle\left\{v\in S\ :\ v\geq 10\right\},\ p\cdot\left(Pr(\geq_\bullet) + \frac{1-Pr(\geq_\bullet)}{2}\right)\cdot\right.&\nonumber\\
  &&&\hspace{5cm}\left.\left(Pr(Rd) + \frac{1-Pr(Rd)}{\mathtt{MAXINT}-\mathtt{MININT}+1}\right)\right\rangle&\label{eq:sp4}&
\end{flalign}

Further, for any program $\sigma,\ sp\big(P = \langle\{\ \}, p\rangle,\ \sigma\big) = \langle\{\ \}, p\rangle$. If the predicate $P$ defines a relation which is not satisfied by any possible program state, or speaking set theoretically if $P$ is an empty set, then it is not possible for $\sigma$ to execute at all since there is no state to start from. Thus the set of states reachable is also empty. Hence, the strongest post-condition is the empty set.

Using the notion of $sp$, we are interested in computing the set of reachable program states for every program point which is known as \emph{collecting semantics}. For each program point $v_i$ (in fig.~\ref{example-program-figure}), let $g_i$ denotes the set of reachable program states. We can consider these collection of program states, $\big(g_i\big)$, as pre-condition predicates for the succeeding program statement. In that case, from the \emph{control flow graph} described in fig.~\ref{example-program-figure}, we can infer that the \emph{collecting semantics}, $\big(g_i\big)$, are linked by $sp$ to give the following system of equation.

\begin{tabular}{lcl}
$g_0$ & = & $\Big\langle\Big\{i\in\mathbb{Z}\ \big|\ \mathtt{MININT}\leq i\leq\mathtt{MAXINT}\Big\},\ 1\Big\rangle$ \\
$g_1$ & = & $\displaystyle sp\big(g_0,\ x =_\bullet 0\big)\bigsqcup_L sp\big(g_2,\ x =_\bullet x +_\bullet 3\big)$ \\
$g_2$ & = & $sp\big(g_1,\ x \leq_\bullet9\big)$ \\
$g_3$ & = & $sp\big(g_1,\ x \geq_\bullet10\big)$ \\
\end{tabular}\\

\noindent We can derive an overall function $\mathcal{F}$, such that
\[ \mathcal{F}\big(g_0,\ g_1,\ g_2,\ g_3\big) = \Big(g_0,\ sp\big(g_0,\ x =_\bullet 0\big)\bigsqcup_L sp\big(g_2,\ x =_\bullet x +_\bullet 3\big),\ sp\big(g_1,\ x \leq_\bullet 9\big),\ sp\big(g_1,\ x \geq_\bullet 10\big)\Big) \]
that maps each $g_i$ to new values of $g_i$ iteratively. The least fixed point of $\mathcal{F}$ is the final solution for the system of equations. We can arrive at the solution by computing the sequence, $\mt{F}^n\big(\bot_L,\ \bot_L,\ \bot_L,\ \bot_L\big)$, starting with the least elements of $\mathcal{L}$ i.e. $\langle\{\ \},1\rangle$. For the sake of the example we assume all the success probabilities to be $1-10^{-4}$, i.e $Pr(+_\bullet)=Pr(-_\bullet)=Pr(Rd)=Pr(Wr)=Pr(\leq_\bullet)=\quad\cdots\quad=1-10^{-4}$ and values of $\mathtt{MININT}$ and $\mathtt{MAXINT}$ to be $-32768$ and $32767$ respectively.
\[ \therefore\quad\left(Pr(+_\bullet)+\frac{1-Pr(+_\bullet)}{\mathtt{MAXINT}-\mathtt{MININT}+1}\right)=0.99990000152=x\text{ (say)},\quad\left(Pr(\leq_\bullet)+\frac{1-Pr(\leq_\bullet)}{2}\right)=0.99995=y\text{ (say)} \]

Solving $\mt{F}$ iteratively would give us a solution as follows :

\noindent
$
\bigg(\bot_L,\ \bot_L,\ \bot_L,\ \bot_L\bigg)\ \to\ 
\bigg(\Big\langle\Big\{a\in\mathbb{Z}\ \big|\ \mathtt{MININT}\leq a\leq\mathtt{MAXINT}\Big\},\ 1\Big\rangle,\ \bullet,\ \bullet,\ \bullet\bigg)\ \to\ 
\bigg(\bullet,\ \Big\langle\Big\{0\Big\},\ x\Big\rangle,\ \bullet,\ \bullet\bigg)\ \to\ \bigg(\bullet,\ \bullet,\ \Big\langle\Big\{0\Big\},\ x^2y\Big\rangle,\ \bullet\bigg)\ \to\ \bigg(\bullet,\ \Big\langle\Big\{0,3\Big\},\ x^5y\Big\rangle,\ \bullet,\ \bullet\bigg)\ \to\ \bigg(\bullet,\ \bullet,\ \Big\langle\Big\{0,3\Big\},\ x^6y^2\Big\rangle,\ \bullet\bigg)\ \to\ \bigg(\bullet,\ \Big\langle\Big\{0,3,6\Big\},\ x^9y^2\Big\rangle,\ \bullet,\ \bullet\bigg)\ \to\ \bigg(\bullet,\ \bullet,\ \Big\langle\Big\{0,3,6\Big\},\ x^{10}y^3\Big\rangle,\ \bullet\bigg)\ \to\ \bigg(\bullet,\ \Big\langle\Big\{0,3,6,9\Big\},\ x^{13}y^3\Big\rangle,\ \bullet,\ \bullet\bigg)\ \to\ \bigg(\bullet,\ \bullet,\ \Big\langle\Big\{0,3,6,9\Big\},\ x^{14}y^4\Big\rangle,\ \bullet\bigg)\ \to\ \bigg(\bullet,\ \Big\langle\Big\{0,3,6,9,12\Big\},\ x^{17}y^4\Big\rangle,\ \bullet,\ \bullet\bigg)\ \ \to\ \bigg(\bullet,\ \bullet,\ \bullet,\ \Big\langle\Big\{12\Big\},\ x^{18}y^5\Big\rangle\bigg)
$\\
       
The `$\bullet$' indicates, repetition of value from previous iteration. The overall solution is therefore

\noindent$\bigg(\Big\langle\Big\{a\in\mathbb{Z}\ \big|\ \mathtt{MININT}\leq a\leq\mathtt{MAXINT}\Big\},\ 1\Big\rangle,\ \Big\langle\Big\{0,3,6,9,12\Big\},\ x^{17}y^4\Big\rangle,\ \Big\langle\Big\{0,3,6,9\Big\},\ x^{14}y^4\Big\rangle,\ \Big\langle\Big\{12\Big\},\ x^{18}y^5\Big\rangle\bigg)\ =$

\noindent$\bigg(\Big\langle\Big\{a\in\mathbb{Z}\ \big|\ -32768\leq a\leq 32767\Big\},\ 1\Big\rangle,\ \Big\langle\Big\{0,3,6,9,12\Big\},\ 0.99810173981\Big\rangle,\ \Big\langle\Big\{0,3,6,9\Big\},\ 0.99840122568\Big\rangle,\\
\Big\langle\Big\{12\Big\},\ 0.99795203106\Big\rangle\bigg)$

In general, we may need infinite number of iterations to converge. This is because the height of the lattice $\mt{L}$ is $\infty$. But for practical purpose we stop iterating as soon as the values converge. On the line of abstract interpretation, the next step is to define a suitable abstract domain for the program states. For each program state in the concrete domain, there exists a corresponding abstract state which approximates the information contained in the concrete program state. An abstract domain is a collection of all such abstract program states. In the next section we define an interval abstract domain corresponding to the concrete domain $(\mt{L}, \sqsubseteq_L)$ and call it the \emph{probabilistic interval abstract domain}.

\section{The Probabilistic Interval Abstract Domain, $(M,\sqsubseteq_M)$}

As in the case with probabilistic concrete domain, the idea behind probabilistic interval domain is also an extension of the standard interval domain of programs. Given a program  $\sigma$, its interval abstract domain implies ranges (unlike set of discrete values as in the case with concrete domain) over its program states at each program point of $\sigma$. Similarly a probabilistic interval domain describes range of values within which the program variables can assume its values along with the probabilities of assuming variable values from the corresponding range. We represent a probabilistic interval domain as a tuple of range of values and its associated probability. This is the minimum probability with which a program variable will take its values from the range. For program with a single integer variable, we define the probabilistic interval abstract domain, $M\subseteq\mathbb{Z}^2\times\mathbb{R}$, as follows.
\begin{definition} Let $\langle [a,b], p_{ab}\rangle$ denote that at a particular program point, a program variable takes its value  within the interval $[a,b]$ with probability $p_{ab}$. $M=\left\{\langle[a,b],\ p_{ab} \rangle\ \middle|\ \mathtt{MININT}\leq a\leq b\leq\mathtt{MAXINT} \bigwedge 0\leq p_{ab}\leq1\right\}$.  In general, for programs with $n$ integer variables, the concrete domain will be $M^n$.
\end{definition}
Before defining a partial order relation on $M$, we define a utility function $\mathit{p.m.f}$ (probability mass function of an interval) as follows:
\begin{definition}\label{pmf}
$\displaystyle p.m.f\Big(\langle[a,b],p_{ab}\rangle\Big)=\frac{p_{ab}}{b-a+1}$
\end{definition}
Using the conventional partial order relation on intervals i.e,
\begin{definition}\label{interval_def}
$[a,\ b]\sqsubseteq_{int}[c,\ d]\iff (c\leq a)\bigwedge\ (b\leq d)$
\end{definition}
we define a partial order $\sqsubseteq_M$ on $M$ as follows:
\begin{definition}\label{abstract_order}
$\forall \langle[a,b],p_{ab}\rangle, \langle[c,d],p_{cd}\rangle\in M\\
\langle[a,b],p_{ab}\rangle\sqsubseteq_M\langle[c,d],p_{cd}\rangle\quad\iff\quad [a,b]\sqsubseteq_{int}[c,d]\ \bigwedge\ p.m.f\Big(\langle[a,b],p_{ab}\rangle\Big)\geq\ p.m.f\Big(\langle[c,d],p_{cd}\rangle\Big)$
\end{definition}
The intuition behind the definition of $\sqsubseteq_M$ can be explained as follows. The element $\langle[c,d],p_{cd}\rangle$ which is higher in the order is an approximation of $\langle[a,b],p_{ab}\rangle$ such that the element $\langle[a,b],p_{ab}\rangle$ provides more precise information about program state since a variables value is predicted to be in a smaller range with higher probability mass.

\begin{lemma}
$(M,\ \sqsubseteq_M)$ is a \textbf{poset}.
\begin{proof}
We prove that $\sqsubseteq_M$ is a partial order relation by showing reflexivity, transitivity and anti-symmetricity for $\sqsubseteq_M$. 
\begin{description}
\item[Reflexivity :] It's trivial to show that $\langle[a,b],p_{ab}\rangle\sqsubseteq_M\langle[a,b],p_{ab}\rangle$. It follows directly from the def.~\ref{abstract_order}.
\item[Transitivity :] Given $\langle[a,b],p_{ab}\rangle\sqsubseteq_M\langle[c,d],p_{cd}\rangle\quad\mathrm{and}\quad\langle[c,d],p_{cd}\rangle\sqsubseteq_M\langle[e,f],p_{ef}\rangle$, we have,

$[a,b]\sqsubseteq_{int}[c,d] \ \bigwedge\ [c,d]\sqsubseteq_{int}[e,f]\quad\implies\quad[a,b]\sqsubseteq_{int}[c,d]$

$p.m.f\Big(\langle[a,b],p_{ab}\rangle\Big)\geq\ p.m.f\Big(\langle[c,d],p_{cd}\rangle\Big)\ \bigwedge\ p.m.f\Big(\langle[c,d],p_{cd}\rangle\Big)\geq\ p.m.f\Big(\langle[e,f],p_{ef}\rangle\Big)$

$\implies p.m.f\Big(\langle[a,b],p_{ab}\rangle\Big)\geq\ p.m.f\Big(\langle[e,f],p_{ef}\rangle\Big)$

Hence, $\langle[a,b],p_{ab}\rangle\sqsubseteq_M\langle[e,f],p_{ef}\rangle$
\item[Anti-Symmetricity :] Given $\langle[a,b],p_{ab}\rangle\sqsubseteq_M\langle[c,d],p_{cd}\rangle\quad\mathrm{and}\quad\langle[c,d],p_{cd}\rangle\sqsubseteq_M\langle[a,b],p_{ab}\rangle$, we have,

$[a,b]\sqsubseteq_{int}[c,d] \ \bigwedge\ [c,d]\sqsubseteq_{int}[a,b]\quad\implies\quad[a,b]=[c,d]$

$p.m.f\Big(\langle[a,b],p_{ab}\rangle\Big)\geq\ p.m.f\Big(\langle[c,d],p_{cd}\rangle\Big)\ \bigwedge\ p.m.f\Big(\langle[c,d],p_{cd}\rangle\Big)\geq\ p.m.f\Big(\langle[a,b],p_{ab}\rangle\Big)$

$\implies p.m.f\Big(\langle[a,b],p_{ab}\rangle\Big) = p.m.f\Big(\langle[c,d],p_{cd}\rangle\Big)$

$\displaystyle\implies \frac{p_{ab}}{b-a+1} = \frac{p_{cd}}{d-c+1} \implies p_{ab} = p_{cd}$\\

Hence, $\langle[a,b],p_{ab}\rangle\ = \langle[c,d],p_{cd}\rangle$
\end{description}
\end{proof}
\end{lemma}

\begin{lemma}
$(M,\ \sqsubseteq_M)$ is a \textbf{complete lattice}.
\begin{proof}
For proving $(M, \sqsubseteq_M)$ to be a \textbf{lattice}, we need to show that for any two elements in $M$, there exist a unique infimum (greatest lower bound) and a unique supremum (least upper bound). The least upper bound $(l.u.b)$ operator, $\displaystyle\bigsqcup_M$ and the greatest lower bound $(g.l.b)$ operator,  $\displaystyle\bigsqcap^M$ for any two elements $\langle [a,b],p_{ab}\rangle, \langle [c,d],p_{cd}\rangle \in M$ can be defined as follows.

\begin{equation}
\label{eq:lub_M}
 \left.\begin{aligned}
        \langle[a,b],p_{ab}\rangle\ \bigsqcup_M\ \langle[\ ],p_\phi\rangle&&=&\quad \langle[a,b],p_{ab}\rangle\\
        \langle[a,b],p_{ab}\rangle\ \bigsqcup_M\ \langle[c,d],p_{cd}\rangle&&=&\quad\ \ \langle[x,y],p_{xy}\rangle\qquad\quad\qquad\qquad
       \end{aligned}\qquad
 \right\}
\end{equation}
where,
	$\quad x=\mathbf{min}(a,c)$,
	$\quad y=\mathbf{max}(b,d)\quad$ and\\
	$\displaystyle p_{xy}=(y-x+1)\ \cdot\ \mathbf{min}\left(p.m.f\Big(\big<[a,\ b],\ p_{ab}\big>\Big),\ p.m.f\Big(\big<[c,\ d],\ p_{cd}\big>\Big),\ \frac{1}{y-x+1}\right)$
	
\begin{equation}
\label{eq:glb_M}
 \left.\begin{aligned}
        \langle[a,b],p_{ab}\rangle\ \bigsqcap^M\ \langle[\ ],p_\phi\rangle&&=&\quad\ \ \langle[\ ],p_\phi\rangle&\\
        \langle[a,b],p_{ab}\rangle\ \bigsqcap^M\ \langle[c,d],p_{cd}\rangle&&=&\begin{cases} 
   			  \ \ \langle[x,y],p_{xy}\rangle&\qquad\qquad\text{if }x\leq y\\
   			  \ \ \langle[\ ],1\rangle&\qquad\qquad\text{otherwise}
  			\end{cases}
       \end{aligned}\quad
 \right\}
\end{equation}
where,
	$\quad x=\mathbf{max}(a,c)$,
	$\quad y=\mathbf{min}(b,d)\quad$ and\\
	$\displaystyle p_{xy}=(y-x+1)\ \cdot\ \mathbf{max}\left(p.m.f\Big(\big<[a,\ b],\ p_{ab}\big>\Big),\ p.m.f\Big(\big<[c,\ d],\ p_{cd}\big>\Big)\right)$
	
Soundness of these definitions can be established as follows. Let's consider the following 3 cases.

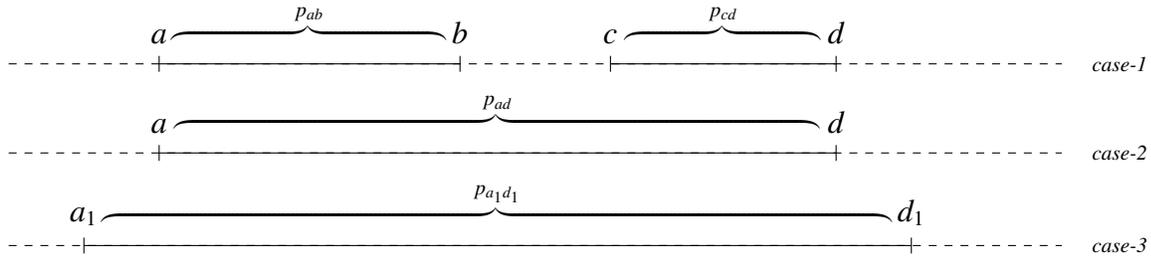
\usetikzlibrary{arrows}
\noindent\begin{tikzpicture}[font=\large]
\draw [dashed] (0,0) -- (14,0);
\foreach \x in {2,6,8,11}
\draw [shift={(\x,0)},color=black] (0,3pt) -- (0,-3pt);
\draw (2,0) -- (6,0);
\draw (8,0) -- (11,0);
\draw (2,0) node[above=4pt]{$a$};
\draw (6,0) node[above=4pt]{$b$};
\draw (8,0) node[above=4pt]{$c$};
\draw (11,0) node[above=4pt]{$d$};
\draw (4,0) node[above=4pt]{$\overbrace{\hspace{105pt}}^{p_{ab}}$};
\draw (9.5,0) node[above=4pt]{$\overbrace{\hspace{75pt}}^{p_{cd}}$};
\draw (14,0) node[right=8pt]{\small{\textit{case-1}}};
\end{tikzpicture}

\noindent\begin{tikzpicture}[font=\large]
\draw [dashed] (0,0) -- (14,0);
\foreach \x in {2,11}
\draw [shift={(\x,0)},color=black] (0,3pt) -- (0,-3pt);
\draw (2,0) -- (11,0);
\draw (2,0) node[above=4pt]{$a$};
\draw (11,0) node[above=4pt]{$d$};
\draw (6.5,0) node[above=4pt]{$\overbrace{\hspace{245pt}}^{p_{ad}}$};
\draw (14,0) node[right=8pt]{\small{\textit{case-2}}};
\end{tikzpicture}

\noindent\begin{tikzpicture}[font=\large]
\draw [dashed] (0,0) -- (14,0);
\foreach \x in {1,12}
\draw [shift={(\x,0)},color=black] (0,3pt) -- (0,-3pt);
\draw (1,0) -- (12,0);
\draw (1,0) node[above=4pt]{$a_1$};
\draw (12,0) node[above=4pt]{$d_1$};
\draw (6.5,0) node[above=4pt]{$\overbrace{\hspace{300pt}}^{p_{a_1d_1}}$};
\draw (14,0) node[right=8pt]{\small{\textit{case-3}}};
\end{tikzpicture}\\

For the two elements $\langle[a,b],p_{ab}\rangle$ and $\langle[c,d],p_{cd}\rangle$ (in {\small\textit{case-1}}) suppose the $l.u.b$ is $\langle[x,y],p_{xy}\rangle$, i.e
\[ \displaystyle\langle[a,b],p_{ab}\rangle\ \bigsqcup_M\ \langle[c,d],p_{cd}\rangle=\langle[x,y],p_{xy}\rangle \]
From eq.~\ref{eq:lub_M} we have,
\[ x=\mathbf{min}(a,c)=a,\quad y=\mathbf{max}(b,d)=d,\quad\displaystyle p_{xy}=(d-a+1)\ \cdot\ \mathbf{min}\left(p.m.f\Big(\big<[a,\ b],\ p_{ab}\big>\Big),\ p.m.f\Big(\big<[c,\ d],\ p_{cd}\big>\Big),\ \frac{1}{d-a+1}\right) \]
$\therefore\langle[a,d],p_{ad}\rangle$ (in {\small\textit{case-2}}) is the $l.u.b$ of the elements $\langle[a,b],p_{ab}\rangle$ and $\langle[c,d],p_{cd}\rangle$ (in {\small\textit{case-1}}) with $p_{ad}=p_{xy}$. Now we show that all upper bounds of $\langle[a,b],p_{ab}\rangle$ and $\langle[c,d],p_{cd}\rangle$ come higher in the order than $\langle[a,d],p_{ad}\rangle$.

It's trivial to show that for any tuple of the form $\langle[a,d],p'_{ad}\rangle$, where $p'_{ad}\leq p_{ad}\footnote{If $p'_{ad}>p_{ad}$, then $\langle[a,d],p'_{ad}\rangle$ is \textbf{NOT} an upper bound of both $\langle[a,b],p_{ab}\rangle$ and $\langle[c,d],p_{cd}\rangle$}$, $\langle[a,d],p_{ad}\rangle\sqsubseteq_M\langle[a,d],p'_{ad}\rangle$ (by def.~\ref{abstract_order}). So, $\langle[a,d],p_{ad}\rangle$ remains the $l.u.b$.

Now, the other form of tuple i.e $\langle[a_1,d_1],p_{a_1d_1}\rangle$ (in {\small\textit{case-3}}), where $[a,d]\sqsubseteq_{int}[a_1,d_1]$, will be an upper bound of $\langle[a,b],p_{ab}\rangle$ and $\langle[c,d],p_{cd}\rangle$ if\\
$\langle[a,b],p_{ab}\rangle\sqsubseteq_M\langle[a_1,d_1],p_{a_1d_1}\rangle\quad\bigwedge\quad\langle[c,d],p_{cd}\rangle\sqsubseteq_M\langle[a_1,d_1],p_{a_1d_1}\rangle$\\
$\Rightarrow p.m.f\Big(\langle[a_1,d_1],p_{a_1d_1}\rangle\Big)\leq p.m.f\Big(\langle[a,b],p_{ab}\rangle\Big)\ \bigwedge\ p.m.f\Big(\langle[a_1,d_1],p_{a_1d_1}\rangle\Big)\leq p.m.f\Big(\langle[c,d],p_{cd}\rangle\Big)$\\
$\Rightarrow p_{a_1d_1}\leq(d_1-a_1+1)\cdot p.m.f\Big(\langle[a,b],p_{ab}\rangle\Big)\ \bigwedge\ p_{a_1d_1}\leq(d_1-a_1+1)\cdot p.m.f\Big(\langle[c,d],p_{cd}\rangle\Big)\hfill\Big[\text{from\ def.}~\ref{pmf}\Big]$\\
$\Rightarrow p_{a_1d_1}\leq\mathbf{min}\left((d_1-a_1+1)\cdot p.m.f\Big(\langle[a,b],p_{ab}\rangle\Big),\ (d_1-a_1+1)\cdot p.m.f\Big(\langle[c,d],p_{cd}\rangle\Big),\ 1\right)\hfill\Big[\because p_{a_1d_1}\leq1\Big]$\\
$\Rightarrow\displaystyle p_{a_1d_1}\leq(d_1-a_1+1)\ \cdot\ \mathbf{min}\left(p.m.f\Big(\big<[a,\ b],\ p_{ab}\big>\Big),\ p.m.f\Big(\big<[c,\ d],\ p_{cd}\big>\Big),\ \frac{1}{d_1-a_1+1}\right)$

\noindent We need to prove that, $p.m.f\Big(\big<[a,\ d],\ p_{ad}\big>\Big)\geq p.m.f\Big(\big<[a_1,\ d_1],\ p_{a_1d_1}\big>\Big)$, in order to prove that $\big<[a,\ d],\ p_{ad}\big>$ still remains the $l.u.b$. Without any loss of generality, we assume that
\begin{alignat}{2}
p.m.f\Big(\langle[a,b],p_{ab}\rangle\Big)\leq p.m.f\Big(\langle[c,d],p_{cd}\rangle\Big)\label{eq:ub1}\\
\frac{1}{d_1-a_1+1}<\frac{1}{d-a+1}\label{eq:ub2}
\end{alignat}
\begin{equation*}
 \left.\begin{aligned}
        p_{ad}&=(d-a+1)\ \cdot\ \mathbf{min}\left(p.m.f\Big(\langle[a,b],p_{ab}\rangle\Big),\ p.m.f\Big(\langle[c,d],p_{cd}\rangle\Big),\ \frac{1}{d-a+1}\right)\\
        &=(d-a+1)\ \cdot\ \mathbf{min}\left(p.m.f\Big(\langle[a,b],p_{ab}\rangle\Big),\ \frac{1}{d-a+1}\right)\qquad\qquad\qquad[\text{from eq.}~\ref{eq:ub1}]\\
        p_{a_1d_1}&=(d_1-a_1+1)\ \cdot\ \mathbf{min}\left(p.m.f\Big(\langle[a,b],p_{ab}\rangle\Big),\ p.m.f\Big(\langle[c,d],p_{cd}\rangle\Big),\ \frac{1}{d_1-a_1+1}\right)\\
        &=(d_1-a_1+1)\ \cdot\ \mathbf{min}\left(p.m.f\Big(\langle[a,b],p_{ab}\rangle\Big),\ \frac{1}{d_1-a_1+1}\right)\qquad\qquad[\text{from eq.}~\ref{eq:ub1}]
       \end{aligned}\qquad
 \right\}\hfill\text{(Given)}
\end{equation*}

In view of eq.~\ref{eq:ub2} there are three cases to consider now.

\begin{description}
	\item[Case 1 : $\qquad\displaystyle p.m.f\Big(\langle\lbrack a,b\rbrack,p_{ab}\rangle\Big)\leq\frac{1}{d_1-a_1+1}$] \hfill
	\[ \therefore \displaystyle p.m.f\Big(\langle[a,b],p_{ab}\rangle\Big)<\frac{1}{d-a+1}\hfill\text{[from eq.~\ref{eq:ub2}]}\]
	\[ \therefore p_{ad}=(d-a+1)\ \cdot\ p.m.f\Big(\langle[a,b],p_{ab}\rangle\Big),\qquad p_{a_1d_1}=(d_1-a_1+1)\ \cdot\ p.m.f\Big(\langle[a,b],p_{ab}\rangle\Big) \]
	\[ \therefore p.m.f\Big(\langle[a,d],p_{ad}\rangle\Big)=p.m.f\Big(\langle[a_1,d_1],p_{a_1d_1}\rangle\Big) \]
	\item[Case 2 : $\qquad\displaystyle \frac{1}{d_1-a_1+1}<p.m.f\Big(\langle\lbrack a,b\rbrack,p_{ab}\rangle\Big)<\frac{1}{d-a+1}$] \hfill
	\[ \therefore p_{ad}=(d-a+1)\ \cdot\ p.m.f\Big(\langle[a,b],p_{ab}\rangle\Big),\qquad p_{a_1d_1}=1 \]
	\[ \therefore p.m.f\Big(\langle[a,d],p_{ad}\rangle\Big)=p.m.f\Big(\langle[a,b],p_{ab}\rangle\Big),\qquad p.m.f\Big(\langle[a_1,d_1],p_{a_1d_1}\rangle\Big)=\frac{1}{d_1-a_1+1} \]
	\[ \therefore p.m.f\Big(\langle[a,d],p_{ad}\rangle\Big)>p.m.f\Big(\langle[a_1,d_1],p_{a_1d_1}\rangle\Big) \]
	\item[Case 3 : $\qquad\displaystyle \frac{1}{d-a+1}\leq p.m.f\Big(\langle\lbrack a,b\rbrack,p_{ab}\rangle\Big)$] \hfill
	\[ \therefore \displaystyle p.m.f\Big(\langle[a,b],p_{ab}\rangle\Big)>\frac{1}{d_1-a_1+1}\hfill\text{[from eq.~\ref{eq:ub2}]} \]
	\[ \therefore p_{ad}=1,\qquad p_{a_1d_1}=1 \]
	\[ \therefore p.m.f\Big(\langle[a,d],p_{ad}\rangle\Big)=\ p.m.f\Big(\langle[a_1,d_1],p_{a_1d_1}\rangle\Big) \]
\end{description}

\noindent In all the above three cases $p.m.f\Big(\langle[a,d],p_{ad}\rangle\Big)\geq p.m.f\Big(\langle[a_1,d_1],p_{a_1d_1}\rangle\Big)$. Hence $\langle[a,d],p_{ad}\rangle$ is the least upper bound. Analogous arguments can be applied to show the soundness of the greatest lower bound, $\displaystyle\bigsqcap^M$ definition. The greatest and least element of $(M,\ \sqsubseteq_M)$ are $\top_M$ and $\bot_M$ respectively and they are defined as, 
$\top_M=\langle[\mathtt{MININT},\mathtt{MAXINT}],\ 0\rangle$ and $\bot_M=\langle[\ ], 1\rangle$. Hence, $\displaystyle\left\langle M, \sqsubseteq_M, \bigsqcup_M, \bigsqcap^M \right\rangle$ is a lattice with greatest and least elements $\top_M$ and  $\bot_M$ respectively. The operations $\displaystyle\bigsqcup_M \text{and} \bigsqcap^M$ can be extended over any $\mt{S} = \{\langle [a_1,b_1], p_1\rangle, \cdots , \langle [a_n,b_n], p_n\rangle\}\subseteq M$ such that $\mt{S}$ has a unique $l.u.b$ and $g.l.b$ in $M$ given by \[ \displaystyle\bigsqcup_M \mt{S} = \left\langle [min_a,\ max_b],\ (max_b-min_a+1)\cdot\textbf{min}\left(\frac{p_1}{b_1-a_1+1},\cdots,\frac{p_n}{b_n-a_n+1},\frac{1}{max_b-min_a+1}\right) \right\rangle \]
with, $min_a=\textbf{min}(a_1,\cdots,a_n),\  max_b=\textbf{max}(b_1,\cdots,b_n)$ and \[ \displaystyle\bigsqcap^M \mt{S} = \left\langle [max_a,\ min_b],\ (min_b-max_a+1)\cdot\textbf{max}\left(\frac{p_1}{b_1-a_1+1},\cdots,\frac{p_n}{b_n-a_n+1},\frac{1}{min_b-max_a+1}\right) \right\rangle \]
with, $max_a=\textbf{max}(a_1,\cdots,a_n),\  min_b=\textbf{min}(b_1,\cdots,b_n)$ respectively. Hence $M$ is a complete lattice.
\end{proof}	
\end{lemma}

\subsection{Hasse Diagram of Lattice $(M,\ \sqsubseteq_M)$}

\begin{figure}
  \resizebox{\linewidth}{!}{
    \input{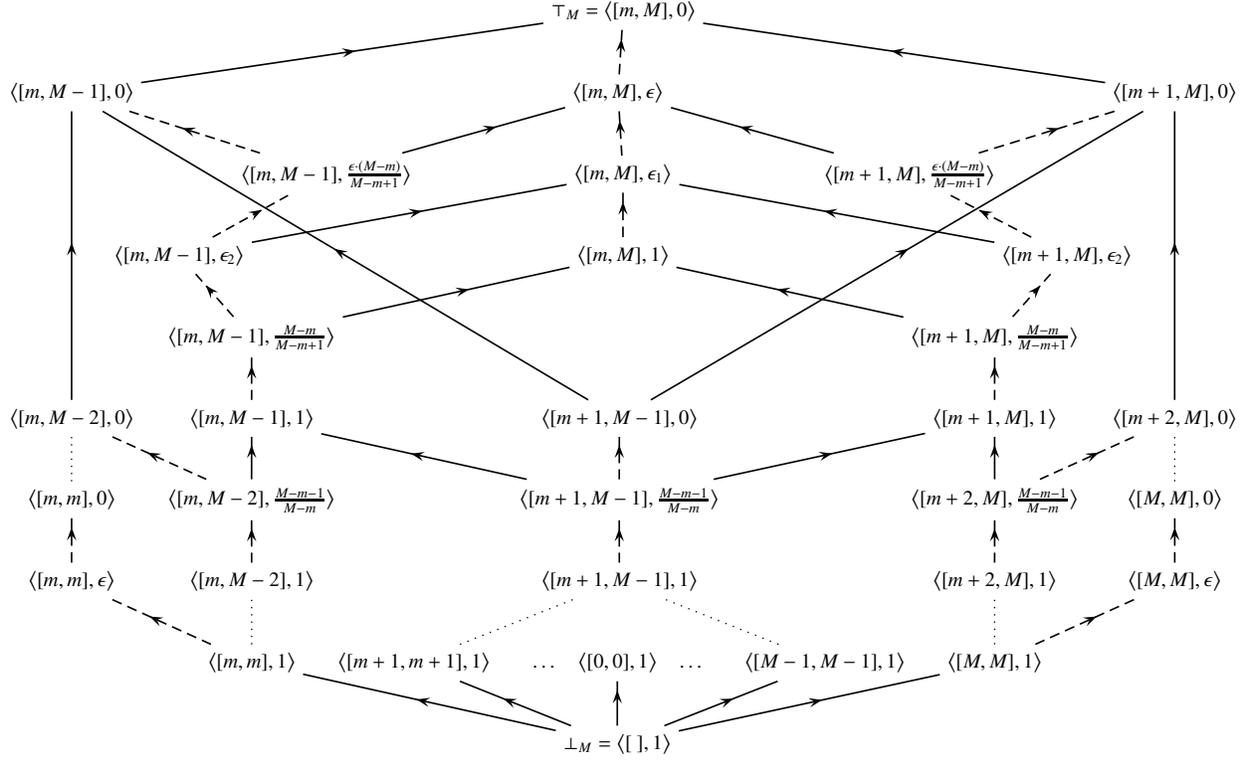}
  }
  \caption{2D Projectional View of $(M,\sqsubseteq_M)$}
  \label{abstract-2d-figure}
\end{figure}

This lattice, $(M,\sqsubseteq_M)$, is structurally similar to that of the concrete one. It is best viewed in 3 dimensions. For that reason, we've given several different projectional views of the same lattice. In all of the projectional views of the abstract lattice, \emph{directed solid lines} represent direct parent-child relationship i.e there doesn't exist any other element between the connecting nodes and \emph{directed dashed lines} represent the existence of infinite number of elements having same intervals as that of the connecting nodes but probability varies monotonically between the connecting nodes.

\begin{figure}
  \centering
  \resizebox{\linewidth}{!}{
    \begin{tikzpicture}[every node/.style={draw, circle, inner sep=0pt, outer sep=0pt, font=\small}, scale=0.5]
      \input{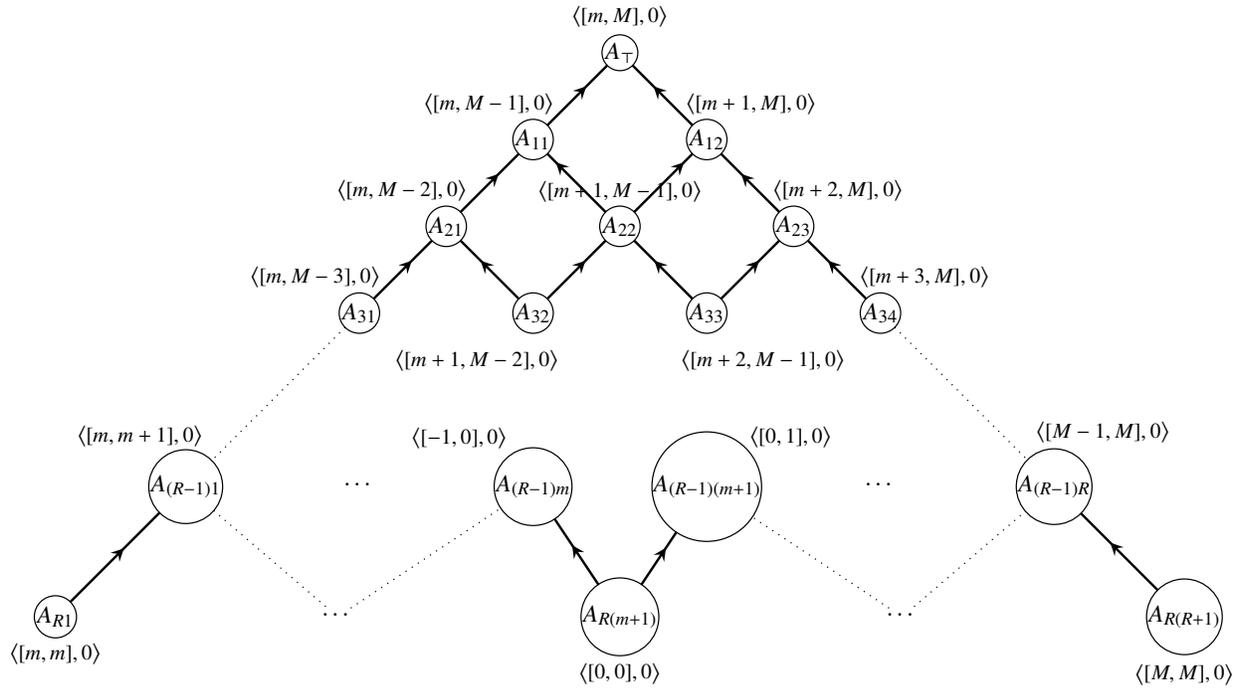}
    \end{tikzpicture}
  }
  \caption{Frontal View of $(M,\sqsubseteq_M)$}
  \label{abstract-frontal-figure}
\end{figure}

In fig.~\ref{abstract-2d-figure}, fig.~\ref{abstract-frontal-figure}, fig.~\ref{abstract-lateral-figure} and fig.~\ref{abstract-rear-figure}, we show respectively the 2D projectional view, the frontal view, a skewed lateral view and the rear view of the abstract lattice.

\begin{description}
  \item[In fig.~\ref{abstract-2d-figure} (2D Projectional View):] \hfill
	\begin{itemize}
	  \item $0<\epsilon,\epsilon_1,\epsilon_2<1$
	  \item $\epsilon_2=\displaystyle\frac{\epsilon_1\cdot(M-m)}{M-m+1}\ \ \footnote{In general, probability of an element = $\displaystyle\frac{\mathrm{(Probability\ of\ its\ direct\ parent)}\times\mathrm{(No.\ of\ elements\ in\ its\ own\ interval)}}{\mathrm{(No.\ of\ elements\ in\ its\ direct\ parent's\ interval)}}$}$
	\end{itemize}
  \item[In fig.~\ref{abstract-2d-figure}, \ref{abstract-frontal-figure}, \ref{abstract-lateral-figure}, \ref{abstract-rear-figure}:] \hfill
    \begin{itemize}
      \item $m=\mathtt{MININT}$
	  \item $M=\mathtt{MAXINT}$
      \item $A_{Ri}\equiv A_{(M-m)i}\quad\text{and}\quad\hat{A}_{Ri}\equiv\hat{A}_{(M-m)i}\qquad\forall i$
      \item For every pair of indices $i, j$ (and $\top$) the elements denoted by $A_{ij}$ (and $A_\top$) and $\hat{A}_{ij}$ (and $\hat{A}_\top$) have the same interval, but probability in element $A_{ij}$ (and $A_\top$) is $0$ , where as it is 1 for $\hat{A}_{ij}$ (and $\hat{A}_\top$).\\
    E.g: $A_{22}\equiv\big<[m+1,\ M-1],\ 0\big>$ but $\hat{A}_{22}\equiv\big<[m+1,\ M-1],\ 1\big>$
    \end{itemize}
\end{description}

\begin{figure}
  \centering
  \resizebox{\linewidth}{!}{
    \begin{tikzpicture}[x={(6mm,-3.5mm)}, y={(0mm,15mm)}, z={(14mm,0mm)}, every node/.style={inner sep=2pt}, scale=0.8]
      \input{figures/abstract_lateral.tex}
    \end{tikzpicture}
  }
  \caption{Lateral View (\textit{skewed}) of $(M,\sqsubseteq_M)$}
  \label{abstract-lateral-figure}

  \resizebox{0.85\linewidth}{!}{
    \begin{tikzpicture}[scale=1]
      \input{figures/abstract_rear.tex}
    \end{tikzpicture}
  }
  \caption{Rear View of $(M,\sqsubseteq_M)$}
  \label{abstract-rear-figure}
\end{figure}

Now that we've defined both the concrete and the abstract domains, we need to establish some form of connections between them in order for abstract interpretation to work. In the following section we define one such connection, called \emph{Galois Connection}, between $(L, \sqsubseteq_L)$ and $(M, \sqsubseteq_M)$. This is essentially a bidirectional mapping with some certain properties between the two domains.

\section{Galois Connection, $\left\langle L,\sqsubseteq_L\right\rangle \xrightleftharpoons[\gamma]{\alpha} \left\langle M,\sqsubseteq_M\right\rangle$}

For two complete lattices $(L,\sqsubseteq_L)$ and $(M,\sqsubseteq_M)$ a pair $(\ALPHA,\GAMMA)$ of monotonic functions $\ALPHA$$:L\to M\ \text{and}\ $$\GAMMA$$:M\to L$ is called a \textbf{\emph{Galois connection}} if
\[ \forall l\in L : l\sqsubseteq_L \GAMMA\Big(\ALPHA\big(l\big)\Big)\qquad\text{and}\qquad\forall m\in M : \ALPHA\Big(\GAMMA\big(m\big)\Big)\sqsubseteq_M m \]
Before going on with the definition of $(\ALPHA,\GAMMA)$ we define few utility functions over $L$, the concrete domain.

\begin{definition}
  $\quad\forall\langle S,p\rangle\in L,\text{ such that }\ S\neq\phi$
  \begin{flalign*}
    &\mathcal{V}_m\Big(\langle S,p\rangle\Big)=v_m\qquad\qquad\text{where,}\quad v_m\in S\ \bigwedge\ \forall v\in S\ \Rightarrow\ v_m\leq v&\\
    &\mathcal{V}_M\Big(\langle S,p\rangle\Big)=v_M\qquad\qquad\text{where,}\quad v_M\in S\ \bigwedge\ \forall v\in S\ \Rightarrow\ v_M\geq v&
  \end{flalign*}
\end{definition}

$\therefore\mathcal{V}_m$ and $\mathcal{V}_M$ returns respectively the \textbf{minimum} and \textbf{maximum} of all \textit{values} of any tuple in $L$. Using these two functions, we define $(\ALPHA,\GAMMA)$ as follows :
\begin{definition}\label{alpha}
  \begin{flalign*}
    \ALPHA\big(\langle S,p\rangle\big)= 
      \begin{cases} 
        \big\langle[\ ],\ 1\big\rangle=\perp_M &  \text{if}\quad\big\langle S,p\big\rangle=\big\langle\{\ \},\ 1\big\rangle=\perp_L \\
        \bigg\langle\Big[\mathcal{V}_m\Big(\langle S,p\rangle\Big),\ \mathcal{V}_M\Big(\langle S,p\rangle\Big)\Big],\ \mathbf{min}\bigg(1,\ p\cdot\Big(\mathcal{V}_M\Big(\langle S,p\rangle\Big)-\mathcal{V}_m\Big(\langle S,p\rangle\Big)+1\Big)\bigg)\bigg\rangle & \text{otherwise}
      \end{cases}
  \end{flalign*}
\end{definition}

\begin{definition}\label{gamma}
  \begin{align*}
    &\GAMMA\Big(\big\langle[\ ],\ 1\big\rangle\Big)&=&\quad\Big\langle\big\{\ \big\},\ 1\Big\rangle=\perp_L&\\
    &\GAMMA\Big(\big\langle[a,\ b],\ p_{ab}\big\rangle\Big)&=&\quad\bigg\langle\bigg\{i\in\mathbb{Z}\ :\ a\leq i\leq b\bigg\},\ p.m.f\Big(\langle[a,b],p_{ab}\rangle\Big)\bigg\rangle\qquad\qquad\qquad\qquad&
  \end{align*}
\end{definition}

\begin{lemma}
$\ALPHA$ is a monotonic function.
\begin{proof}
Suppose $\langle S_1,p_1\rangle$ and $\langle S_2,p_2\rangle$ are two elements in $\mt{L}$, with $\langle S_1,p_1\rangle\sqsubseteq_L\langle S_2,p_2\rangle$ and $S_1\neq\phi$\footnote{If $\quad S_1=\phi$, we have $\langle S_1,p_1\rangle=\sqsubseteq_L\langle S_2,p_2\rangle\ \Rightarrow\ $$\ALPHA$$\Big(\langle S_1,p_1\rangle\Big)=\bot_M\sqsubseteq_M\ $$\ALPHA$$\Big(\langle S_2,p_2\rangle\Big)$}. We show that $\ALPHA$$\Big(\langle S_1,p_1\rangle\Big)\sqsubseteq_M\ $$\ALPHA$$\Big(\langle S_2,p_2\rangle\Big)$.

\noindent$\because\langle S_1,p_1\rangle\sqsubseteq_L\langle S_2,p_2\rangle$, from def.~\ref{concrete_order}, we have, $S_1\subseteq S_2$ and $p_1\geq p_2$.
\begin{flalign}
&\quad S_1\subseteq S_2&&\nonumber\\
\Rightarrow&\quad\mathcal{V}_m\Big(\langle S_2,p_2\rangle\Big)\leq\mathcal{V}_m\Big(\langle S_1,p_1\rangle\Big)\ \bigwedge\ \mathcal{V}_M\Big(\langle S_1,p_1\rangle\Big)\leq\mathcal{V}_M\Big(\langle S_2,p_2\rangle\Big)&&\nonumber\\
\Rightarrow&\quad\Big[\mathcal{V}_m\Big(\langle S_1,p_1\rangle\Big),\ \mathcal{V}_M\Big(\langle S_1,p_1\rangle\Big)\Big]\ \sqsubseteq_{int}\ \Big[\mathcal{V}_m\Big(\langle S_2,p_2\rangle\Big),\ \mathcal{V}_M\Big(\langle S_2,p_2\rangle\Big)\Big]&&\label{eq:monotone5}\\
\Rightarrow&\quad\mathcal{V}_M\Big(\langle S_1,p_1\rangle\Big)-\mathcal{V}_m\Big(\langle S_1,p_1\rangle\Big)+1\leq\mathcal{V}_M\Big(\langle S_2,p_2\rangle\Big)-\mathcal{V}_m\Big(\langle S_2,p_2\rangle\Big)+1&&\nonumber\\
\Rightarrow&\quad\displaystyle\frac{1}{\mathcal{V}_M\Big(\langle S_1,p_1\rangle\Big)-\mathcal{V}_m\Big(\langle S_1,p_1\rangle\Big)+1}\geq\displaystyle\frac{1}{\mathcal{V}_M\Big(\langle S_2,p_2\rangle\Big)-\mathcal{V}_m\Big(\langle S_2,p_2\rangle\Big)+1}&&\label{eq:monotone6}
\end{flalign}

\noindent$\ALPHA$$\big(\langle S_1,p_1\rangle\big) = \bigg\langle\Big[\mathcal{V}_m\Big(\langle S_1,p_1\rangle\Big),\ \mathcal{V}_M\Big(\langle S_1,p_1\rangle\Big)\Big],\ \mathbf{min}\bigg(1,\ p_1\cdot\Big(\mathcal{V}_M\Big(\langle S_1,p_1\rangle\Big)-\mathcal{V}_m\Big(\langle S_1,p_1\rangle\Big)+1\Big)\bigg)\bigg\rangle\qquad$ and

\noindent$\ALPHA$$\big(\langle S_2,p_2\rangle\big) = \bigg\langle\Big[\mathcal{V}_m\Big(\langle S_2,p_2\rangle\Big),\ \mathcal{V}_M\Big(\langle S_2,p_2\rangle\Big)\Big],\ \mathbf{min}\bigg(1,\ p_2\cdot\Big(\mathcal{V}_M\Big(\langle S_2,p_2\rangle\Big)-\mathcal{V}_m\Big(\langle S_2,p_2\rangle\Big)+1\Big)\bigg)\bigg\rangle$
  
\noindent$\displaystyle p.m.f\Big($$\ALPHA$$\big(\langle S_i,p_i\rangle\big)\Big) = \frac{\mathbf{min}\left(1,\ p_i\cdot\Big(\mathcal{V}_M\big(\langle S_i,p_i\rangle\big)-\mathcal{V}_m\big(\langle S_i,p_i\rangle\big)+1\Big)\right)}{\mathcal{V}_M\Big(\langle S_i,\ p_i\rangle\Big)-\mathcal{V}_m\Big(\langle S_i,p_i\rangle\Big)+1} = \mathbf{min}\left(p_i,\displaystyle\frac{1}{\mathcal{V}_M\Big(\langle S_i,\ p_i\rangle\Big)-\mathcal{V}_m\Big(\langle S_i,p_i\rangle\Big)+1}\right)\quad\forall i\in\{1,2\}$

\noindent Now, depending upon the values of $p.m.f\Big($$\ALPHA$$\big(\langle S_1,p_1\rangle\big)\Big)$ and $p.m.f\Big($$\ALPHA$$\big(\langle S_2,p_2\rangle\big)\Big),$ we have 4 different cases to consider.

\begin{description}
  \item[Case 1 :] $p.m.f\Big($$\ALPHA$$\big(\langle S_1,p_1\rangle\big)\Big)=p_1\qquad$ and $\qquad p.m.f\Big($$\ALPHA$$\big(\langle S_2,p_2\rangle\big)\Big)=p_2$ \hfill \\
    We already know $p_1\geq p_2\qquad\therefore p.m.f\Big($$\ALPHA$$\big(\langle S_1,p_1\rangle\big)\Big)\geq p.m.f\Big($$\ALPHA$$\big(\langle S_2,p_2\rangle\big)\Big)$
  
  \item[Case 2 :] $p.m.f\Big($$\ALPHA$$\big(\langle S_1,p_1\rangle\big)\Big)=p_1\qquad$ and $\qquad p.m.f\Big($$\ALPHA$$\big(\langle S_2,p_2\rangle\big)\Big)=\displaystyle\frac{1}{\mathcal{V}_M\Big(\langle S_2,p_2\rangle\Big)-\mathcal{V}_m\Big(\langle S_2,p_2\rangle\Big)+1}$ \hfill \\
    
    $\because\ p.m.f\Big($$\ALPHA$$\big(\langle S_2,p_2\rangle\big)\Big)=\displaystyle\frac{1}{\mathcal{V}_M\Big(\langle S_2,p_2\rangle\Big)-\mathcal{V}_m\Big(\langle S_2,p_2\rangle\Big)+1}$
    \begin{flalign*}
      \Rightarrow&\quad p_2\geq\frac{1}{\mathcal{V}_M\Big(\langle S_2,p_2\rangle\Big)-\mathcal{V}_m\Big(\langle S_2,p_2\rangle\Big)+1}&&\\
      \Rightarrow&\quad p_1\geq\frac{1}{\mathcal{V}_M\Big(\langle S_2,p_2\rangle\Big)-\mathcal{V}_m\Big(\langle S_2,p_2\rangle\Big)+1} & \text{[$\because p_1\geq p_2$]} &
    \end{flalign*}
  $\therefore\ p.m.f\Big($$\ALPHA$$\big(\langle S_1,p_1\rangle\big)\Big)\geq p.m.f\Big($$\ALPHA$$\big(\langle S_2,p_2\rangle\big)\Big)$
  
  \item[Case 3 :] $p.m.f\Big($$\ALPHA$$\big(\langle S_1,p_1\rangle\big)\Big)=\displaystyle\frac{1}{\mathcal{V}_M\Big(\langle S_1,p_1\rangle\Big)-\mathcal{V}_m\Big(\langle S_1,p_1\rangle\Big)+1}\qquad$ and $\qquad p.m.f\Big($$\ALPHA$$\big(\langle S_2,p_2\rangle\big)\Big)=p_2$ \hfill \\
  
  $\because\ p.m.f\Big($$\ALPHA$$\big(\langle S_2,p_2\rangle\big)\Big)=p_2$
    \begin{flalign*}
      \Rightarrow&\quad\frac{1}{\mathcal{V}_M\Big(\langle S_2,p_2\rangle\Big)-\mathcal{V}_m\Big(\langle S_2,p_2\rangle\Big)+1}\geq p_2&&\\
      \Rightarrow&\quad\frac{1}{\mathcal{V}_M\Big(\langle S_1,p_1\rangle\Big)-\mathcal{V}_m\Big(\langle S_1,p_1\rangle\Big)+1}\geq p_2 & \text{[from eq.~\ref{eq:monotone6}]} &
    \end{flalign*}
  $\therefore\ p.m.f\Big($$\ALPHA$$\big(\langle S_1,p_1\rangle\big)\Big)\geq p.m.f\Big($$\ALPHA$$\big(\langle S_2,p_2\rangle\big)\Big)$
  
  \item[Case 4 :] $p.m.f\Big($$\ALPHA$$\big(\langle S_i,p_i\rangle\big)\Big)=\displaystyle\frac{1}{\mathcal{V}_M\Big(\langle S_i,p_i\rangle\Big)-\mathcal{V}_m\Big(\langle S_i,p_i\rangle\Big)+1}\qquad\forall i\in\{1,2\}$ \hfill \\
    From eq.~\ref{eq:monotone6}, it's obvious that $\qquad p.m.f\Big($$\ALPHA$$\big(\langle S_1,p_1\rangle\big)\Big)\geq p.m.f\Big($$\ALPHA$$\big(\langle S_2,p_2\rangle\big)\Big)$
\end{description}

\noindent So, in all the 4 cases, $\quad p.m.f\Big($$\ALPHA$$\big(\langle S_1,p_1\rangle\big)\Big)\geq p.m.f\Big($$\ALPHA$$\big(\langle S_2,p_2\rangle\big)\Big)$. This result along with eq.~\ref{eq:monotone5} proves that $\ALPHA$$\Big(\langle S_1,p_1\rangle\Big)\sqsubseteq_M\ $$\ALPHA$$\Big(\langle S_2,p_2\rangle\Big)$. This completes the proof that $\ALPHA$ is monotonic.
\end{proof}
\end{lemma}

\begin{lemma}
$\GAMMA$ is a monotonic function.
\begin{proof}

Let, $\langle[a,b],p_{ab}\rangle$ and $\langle[c,d],p_{cd}\rangle$ be two elements in $M$, with $\langle[a,b],p_{ab}\rangle\sqsubseteq_M\langle[c,d],p_{cd}\rangle$\footnote{If $\quad\perp_M\sqsubseteq_M\langle[a,b],p_{ab}\rangle$, we trivially have $\quad$$\GAMMA$$\Big(\perp_M\Big)\sqsubseteq_L\ $$\GAMMA$$\Big(\langle[a,b],p_{ab}\rangle\Big)\quad$[$\ \because$$\GAMMA$$\Big(\perp_M\Big)=\perp_L$ is the least element in $\mathcal{L}$]}. We show that $\GAMMA$$\Big(\langle[a,b],p_{ab}\rangle\Big)\sqsubseteq_L\ $$\GAMMA$$\Big(\langle[c,d],p_{cd}\rangle\Big)$.

\noindent$\GAMMA$$\Big(\langle[a,b],\ p_{ab}\rangle\Big) = \bigg\langle\Big\{i\in\mathbb{Z}\ \Big|\ a\leq i\leq b\Big\},\ p.m.f\Big(\langle[a,b],\ p_{ab}\rangle\Big)\bigg)\bigg\rangle$ and

\noindent$\GAMMA$$\Big(\langle[c,d],\ p_{cd}\rangle\Big) = \bigg\langle\Big\{i\in\mathbb{Z}\ \Big|\ c\leq i\leq d\Big\},\ p.m.f\Big(\langle[c,d],\ p_{cd}\rangle\Big)\bigg)\bigg\rangle$

\noindent$\because\langle[a,b],p_{ab}\rangle\sqsubseteq_M\langle[c,d],p_{cd}\rangle$, from def.~\ref{abstract_order} we have, $[a,b]\sqsubseteq_{int}[c,d]$ and $p.m.f\big(\langle[a,b],p_{ab}\rangle\big)\geq p.m.f\big(\langle[c,d],p_{cd}\rangle\big)$.
\begin{flalign}
  &\quad[a,b]\sqsubseteq_{int}[c,d]&\nonumber\\
  \Rightarrow &\quad c\leq a\bigwedge b\leq d&\nonumber\\
  \Rightarrow &\quad\{i\in\mathbb{Z}\ |\ a\leq i\leq b\}\subseteq\{i\in\mathbb{Z}\ |\ c\leq i\leq d\}\nonumber&\\
  \Rightarrow &\quad\GAMMA\Big(\langle[a,b],p_{ab}\rangle\Big)\sqsubseteq_L\ \GAMMA\Big(\langle[c,d],p_{cd}\rangle\Big)\nonumber&\hfill[\text{using def.~\ref{concrete_order}}]
\end{flalign}
This completes the proof that $\GAMMA$ is monotonic.
\end{proof}
\end{lemma}

\begin{lemma}
$\ALPHA$ and $\GAMMA$ are \emph{Galois Connections} i.e $\forall l\in L : l\sqsubseteq_L \GAMMA\Big(\ALPHA\big(l\big)\Big)\quad\text{and}\quad\forall m\in M : \ALPHA\Big(\GAMMA\big(m\big)\Big)\sqsubseteq_M m$
\begin{proof}
First we show that $\forall l\in L : l\sqsubseteq_L \GAMMA\Big(\ALPHA\big(l\big)\Big)\qquad\text{i.e}\qquad\forall \langle S,p\rangle\in L, \langle S,p\rangle\sqsubseteq_L\ $$\GAMMA$$\Big($$\ALPHA$$\big(\langle S,p\rangle\big)\Big)$.

\noindent This can be proved by expanding the definitions of $\ALPHA$.

If, $\quad\langle S,p\rangle=\langle\phi,1\rangle=\perp_L,\quad$ we trivially have, $\qquad\langle\{\},1\rangle=\ $$\GAMMA$$\Big($$\ALPHA$$\big(\langle\{\},1\rangle\big)\Big)\quad$ i.e $\quad\langle\{\},1\rangle\sqsubseteq_L\ $$\GAMMA$$\Big($$\ALPHA$$\big(\langle\{\},1\rangle\big)\Big).$

Otherwise, $\quad$$\GAMMA$$\Big($$\ALPHA$$\big(\langle S,p\rangle\big)\Big)=\left\langle\left\{i\in\mathbb{Z}\ \middle|\ \mathcal{V}_m\Big(\langle S,p\rangle\Big)\leq i\leq\mathcal{V}_M\Big(\langle S,p\rangle\Big)\right\},\ \mathbf{min}\left(p,\ \displaystyle\frac{1}{\mathcal{V}_M\Big(\langle S,p\rangle\Big)-\mathcal{V}_m\Big(\langle S,p\rangle\Big)+1}\right)\right\rangle$

and $\quad\langle S,p\rangle\sqsubseteq_L\left\langle\left\{i\in\mathbb{Z}\ \middle|\ \mathcal{V}_m\Big(\langle S,p\rangle\Big)\leq i\leq\mathcal{V}_M\Big(\langle S,p\rangle\Big)\right\},\ \mathbf{min}\left(p,\ \displaystyle\frac{1}{\mathcal{V}_M\Big(\langle S,p\rangle\Big)-\mathcal{V}_m\Big(\langle S,p\rangle\Big)+1}\right)\right\rangle\footnote{Trivially follows from def.~\ref{concrete_order}, the definition of $\sqsubseteq_L.$}$\\

\noindent Similarly, we can prove that $\qquad\forall m\in M\quad$$\ALPHA$$\Big($$\GAMMA$$\big(m\big)\Big)\sqsubseteq_M\ m$

If $\quad m=\langle[\ ],1\rangle=\perp_M,\qquad$ we have $\qquad$$\ALPHA$$\Big($$\GAMMA$$\big(\langle[\ ],1\rangle\big)\Big)=\ \langle[\ ],1\rangle\quad$ i.e $\qquad$$\ALPHA$$\Big($$\GAMMA$$\big(\langle[\ ],1\rangle\big)\Big)\sqsubseteq_M\ \langle[\ ],1\rangle\quad$

Otherwise, $\quad$$\ALPHA$$\bigg($$\GAMMA$$\Big(\langle [a,b],p_{ab}\rangle\Big)\bigg) =\ $$\ALPHA$$\left(\left\langle\left\{i\in\mathbb{Z}\ :\ a\leq i\leq b,\right\},\ \displaystyle\frac{p_{ab}}{b-a+1}\right\rangle\right)=\ \langle[a,b],p_{ab}\rangle$\footnote{$\because p_{ab}\leq 1$}

$\therefore\quad$$\ALPHA$$\Big($$\GAMMA$$\big(m\big)\Big) =\ m\sqsubseteq_M\ m\qquad\forall m\in M$
\end{proof}
\end{lemma}

Hence, we've established the Galois Connection in terms of $\big(\ALPHA,\GAMMA\big)$ between our concrete domain, $(L,\sqsubseteq_L)$ and our abstract domain, $(M,\sqsubseteq_M)$. These are all the tools required by abstract interpretation for efficient abstraction of the concrete program states. Abstract interpretation utilizes the Galois Connection, $(\alpha,\gamma)$, to approximate the strongest post-condition function, $sp$, defined over the concrete domain, $\mt{L}$, to another corresponding function, $sp^\#$, defined over the abstract domain, $M$. This approximation is useful in the sense that, elements in the abstract domain, $M$, are computationally easier to work with. This is not very hard to understand -- since, $sp^\#$, the approximation of $sp$, is going to be defined over $M$, it will only work on intervals, whereas its concrete domain counterpart, $sp$, has to work on concrete sets of discrete values. Intuitively, intervals are much simpler elements to work with compared to the sets of discrete values, in view of the number of computational steps required to perform any mathematical operation on intervals is quite less. As we are approximating the strongest post-condition function, it is inevitable that we are going to lose some precision --- results obtained using $sp^\#$ will be less precise than that obtained using $sp$. This is a trade-off between ease of computation and precision that we need to adapt. But, while we accept the imprecision due to approximation, we must not compromise with safety --- i.e values returned by $sp^\#$ may be imprecise but it must be safe at the same time. We explain it with the following example.
\begin{example}
Let's say that a program variable can take values from a set $\{2, 4, 8 ,12\}$ with probability $p$, at a certain program point. As we've seen earlier, these type of concrete domain elements are obtained from $sp$. Now the approximation of $sp$ i.e $sp^\#$ is supposed to generate an abstraction of the concrete state, $\langle\{2, 4, 8 ,12\},\ p\rangle$. So, it may produce abstract states like $\langle[0,12],\ p_1\rangle$ or $\langle[-2,14],\ p_2\rangle$ or so on, which are imprecise but safe -- but it is not supposed to generate an interval like $[3,12]$, which is not safe, as it fails to include a possible value (2 in this case) actually taken by the program variable during execution.
\end{example}

In the following section we'll show how we are using \emph{Galois Connection} to approximate $sp$ by $sp^\#$ and apply it on our example program from fig.~\ref{example-program-figure}.

\section{Safe Approximation of Strongest Post-Condition Function, $sp$, in the Abstract Domain $M$}

For a program statement $\sigma$, we can think of the strongest post-condition function $sp$ as $sp_\sigma : L\to L$. This is a transformation from a function with two arguments i.e. $sp(P,\sigma)$ to a function $sp_\sigma$ with a single argument $P$ which is an element of $L$ (a collection of program states). Given a $\sigma$, we can construct the function $sp_\sigma$ which maps a collection of concrete program states to another collection of concrete program states. We are interested in \textbf{\textit{safe approximation}} of such functions in our abstract domain $M$. Let $sp^\#_\sigma : M\to M$ be our choice of safe approximation in the abstract domain M. For safe approximation of $sp_\sigma$ by $sp^\#_\sigma$ we require the following to hold:
\[ \forall m\in M,\quad$$\ALPHA$$\Big(sp\big($$\GAMMA$$(m),\ \sigma\big)\Big)\sqsubseteq_M sp^\#\big(m,\ \sigma\big)\quad$ or equivalently $\quad sp\Big($$\GAMMA$$\big(m\big),\ \sigma\Big)\sqsubseteq_L\ $$\GAMMA$$\Big(sp^\#\big(m,\ \sigma\big)\Big) \]

The functions $\ALPHA$ and $\GAMMA$ are defined earlier for the probabilistic interval domain abstraction. In case $sp^\#$\footnote{For notational convenience we'll write $sp^\#$ and $sp^\#_\sigma$ interchangeably} is the \textbf{\textit{most precise safe approximation}}, we can write
\begin{equation}
\ALPHA$$\Big(sp\big($$\GAMMA$$(m),\ \sigma\big)\Big)=sp^\#\big(m,\ \sigma\big)\label{eq:safe-approx}
\end{equation}
Therefore, using $\ALPHA$ and $\GAMMA$ from def.~\ref{alpha} and def.~\ref{gamma} respectively, we can use the above eq.~\ref{eq:safe-approx} to compute $sp^\#$. Before doing that let's define some notational convenience as follows:

$\displaystyle Rel(Rd) \overset{\mathrm{def}}{=\joinrel=} Pr(Rd) + \frac{1-Pr(Rd)}{\mathtt{MAXINT}-\mathtt{MININT}+1}\qquad Rel(Wr) \overset{\mathrm{def}}{=\joinrel=} Pr(Wr) + \frac{1-Pr(Wr)}{\mathtt{MAXINT}-\mathtt{MININT}+1}$\\

$\displaystyle Rel(+_{\bullet}) \overset{\mathrm{def}}{=\joinrel=} Pr(+_{\bullet}) + \frac{1-Pr(+_{\bullet})}{\mathtt{MAXINT}-\mathtt{MININT}+1}\qquad Rel(\leq_{\bullet}) \overset{\mathrm{def}}{=\joinrel=} Pr(\leq_{\bullet}) + \frac{1-Pr(\leq_{\bullet})}{2}$\\

$\displaystyle Rel(\geq_{\bullet}) \overset{\mathrm{def}}{=\joinrel=} Pr(\geq_{\bullet}) + \frac{1-Pr(\geq_{\bullet})}{2}$

\noindent For our example program from fig.~\ref{example-program-figure} we have defined 4 different $sp(sp_\sigma)$ in eq.~\ref{eq:sp1}, eq.~\ref{eq:sp2}, eq.~\ref{eq:sp3} and eq.~\ref{eq:sp4}. For each of these equations we define the corresponding $sp^\#(sp^\#_\sigma)$ using the eq.~\ref{eq:safe-approx} as follows:
\begin{flalign}
  &sp^\#\Big(\big\langle[a,b],p\big\rangle,\ x =_\bullet 0\Big)&=\ &\ALPHA\Big(sp\big(\GAMMA\big(\big\langle[a,b],p\big\rangle\big),\ x =_\bullet 0\big)\Big)&\nonumber\\
  &&=\ &\ALPHA\Big(sp\Big(\left\langle\left\{i\in\mathbb{Z}\ :\ a\leq i\leq b\right\},\ p.m.f\big(\langle[a,b],p\rangle\big)\right\rangle,\ x =_\bullet 0\Big)\Big) & \text{[from def.~\ref{gamma}]}\nonumber\\
  &&=\ &\ALPHA\left(\left\langle\Big\{0\Big\},\ Rel(Wr)\right\rangle\right) & \text{[from eq.~\ref{eq:sp1}]}\nonumber\\
  &&=\ &\left\langle\Big[0,0\Big],\ Rel(Wr)\right\rangle\qquad\qquad\text{[from def.~\ref{alpha}]}& \\
  &sp^\#\Big(\big\langle[a,b],p\big\rangle,\ x =_\bullet x +_\bullet 3\Big)&=\ &\left\langle\Big[a+3,b+3\Big],\ p\cdot Rel(Rd)\cdot Rel(+_{\bullet})\cdot Rel(Wr)\right\rangle&\\
  &sp^\#\Big(\big\langle[a,b],p\big\rangle,\ x \leq_\bullet 9\Big)&=\ &
  \begin{cases} 
   \quad\displaystyle\left\langle\Big[a,b\Big],\ p\cdot Rel(Rd)\cdot Rel(\leq_{\bullet})\right\rangle       & \text{if } b < 9\\
   \quad\displaystyle\left\langle\Big[a,9\Big],\ \frac{p\cdot(9-a+1)}{b-a+1}\cdot Rel(Rd)\cdot Rel(\leq_{\bullet})\right\rangle       & \text{if } a\leq 9\leq b\\
   \quad\bot_M       & \text{if } 9 < a
  \end{cases}&\\
  &sp^\#\Big(\big\langle[a,b],p\big\rangle,\ x \geq_\bullet 10\Big)&=\ &
  \begin{cases} 
   \quad\displaystyle\left\langle\Big[a,b\Big],\ p\cdot Rel(Rd)\cdot Rel(\geq_{\bullet})\right\rangle       & \text{if } 10 < a\\
   \quad\displaystyle\left\langle\Big[10,b\Big],\ \frac{p\cdot(b-10+1)}{b-a+1}\cdot Rel(Rd)\cdot Rel(\geq_{\bullet})\right\rangle       & \text{if } a\leq 10\leq b\\
   \quad\bot_M       & \text{if } b < 10
  \end{cases}&\\
  &sp^\#\Big(\bot_M=\big\langle[\ ],1\big\rangle,\ \sigma\Big)&=\ &\ALPHA\Big(sp\big(\GAMMA\big(\bot_M\big),\ \sigma\big)\Big)&\nonumber\\
  &&=\ &\ALPHA\Big(sp\Big(\bot_L,\ \sigma\Big)\Big)&\nonumber\\
  &&=\ &\ALPHA\left(\bot_L\right)\qquad\qquad[\because\ sp(\bot_L,\ \sigma)=\bot_L,\quad\text{for any program statement }\sigma]&\nonumber\\
  &&=\ &\bot_M&
\end{flalign}

With these definitions of $sp^\#$ over $M$, we'll apply it on the example program from fig.~\ref{example-program-figure} and show how the abstract program states are being modified by $sp^\#$ directly and eventually reaching a fixed program state.

\subsubsection{Fixed point computation of abstract program states using $sp^\#$}

Let us now try the reachability computation in the abstract domain. Let the set of abstract program states at different program points of the program described in fig.~\ref{example-program-figure}, be defined as $g_0^\#,\ g_1^\#,\ g_2^\#,\ g_3^\#$ and the abstract version of the overall function $\mathcal{F}$ be defined as $\mathcal{F}^\#$. Then we shall have,\\

\begin{tabular}{lcl}
$g_0^\#$ & = & $\big\langle[\mathtt{MININT},\ \mathtt{MAXINT}],\ 1\big\rangle$ \\
$g_1^\#$ & = & $\displaystyle sp^\#\big(g_0^\#,\ x =_\bullet 0\big)\bigsqcup_M sp^\#\big(g_2^\#,\ x =_\bullet x +_\bullet 3\big)$ \\
$g_2^\#$ & = & $sp^\#\big(g_1^\#,\ x \leq_\bullet 9\big)$ \\
$g_3^\#$ & = & $sp^\#\big(g_1^\#,\ x \geq_\bullet 10\big)$ \\
\end{tabular}

\begin{multline*}
  \mathcal{F}^\#\Big(g_0^\#,\ g_1^\#,\ g_2^\#,\ g_3^\#\Big)=\Big(\big\langle[\mathtt{MININT},\ \mathtt{MAXINT}],\ 1\big\rangle,\ sp^\#\big(g_0^\#,\ x =_\bullet 0\big)\bigsqcup_M sp^\#\big(g_2^\#,\ x =_\bullet x +_\bullet 3\big),\\
  sp^\#\big(g_1^\#,\ x\leq_\bullet 9\big),\ sp^\#\big(g_1^\#,\ x\geq_\bullet 10\big)\Big)
\end{multline*}

For demonstration purpose we again assume all the success probabilities to be $1-10^{-4}$, i.e $Pr(+_\bullet)=Pr(-_\bullet)=Pr(Rd)=Pr(Wr)=Pr(\leq_\bullet)=\quad\cdots\quad=1-10^{-4}$ and value of $\mathtt{MININT}$ and $\mathtt{MAXINT}$ to be $-32768$ and $32767$ respectively. Therefore, $\displaystyle\quad\left(Pr(+_\bullet)+\frac{1-Pr(+_\bullet)}{\mathtt{MAXINT}-\mathtt{MININT}+1}\right)=0.9999000015=x\text{ (say)}\quad\text{and}\quad\left(Pr(\leq_\bullet)+\frac{1-Pr(\leq_\bullet)}{2}\right)=0.99995=y\text{ (say)}$

The least fixed point of $\mathcal{F^\#}$, i.e $(\mathcal{F^\#})^n(\bot_M, \bot_M, \bot_M, \bot_M)$ will be our final solution. The computation of

\noindent$(\mathcal{F^\#})^n(\bot_M, \bot_M, \bot_M, \bot_M)$ will be as follows

\noindent
$
\bigg(\bot_M,\ \bot_M,\ \bot_M,\ \bot_M\bigg)\ \to\ 
\bigg(\big\langle[\mathtt{MININT},\mathtt{MAXINT}],\ 1\big\rangle,\ \bullet,\ \bullet,\ \bullet\bigg)\ \to\ 
\bigg(\bullet,\ \big\langle[0,0],\ x\big\rangle,\ \bullet,\ \bullet\bigg)\ \to$\\
$ 
\bigg(\bullet,\ \bullet,\ \big\langle[0,0],\ x^2y\big\rangle,\ \bullet\bigg)\ \to\ \bigg(\bullet,\ \big\langle[0,3],\ 1\big\rangle,\ \bullet,\ \bullet\bigg)\ \to\ \bigg(\bullet,\ \bullet,\ \big\langle[0,3],\ xy\big\rangle,\ \bullet\bigg)\ \to\ \bigg(\bullet,\ \big\langle[0,6],\ 1\big\rangle,\ \bullet,\ \bullet\bigg)\ \to\ \bigg(\bullet,\ \bullet,\ \big\langle[0,6],\ xy\big\rangle,\ \bullet\bigg)\ \to\ \bigg(\bullet,\ \big\langle[0,9],\ 1\big\rangle,\ \bullet,\ \bullet\bigg)\ \to\ \bigg(\bullet,\ \bullet,\ \big\langle[0,9],\ xy\big\rangle,\ \bullet\bigg)\ \to\ \bigg(\bullet,\ \big\langle[0,12],\ 1\big\rangle,\ \bullet,\ \bullet\bigg)\ \to\ \bigg(\bullet,\ \bullet,\ \bullet,\ \big\langle[10,12],\ \frac{3}{13}xy\big\rangle\bigg)
$\\

The `$\bullet$' indicates, repetition of value from previous iteration. The overall solution is therefore

\noindent$\bigg(\Big\langle[\mathtt{MININT},\leq\mathtt{MAXINT}],\ 1\Big\rangle,\ \Big\langle[0,12],\ 1\Big\rangle,\ \Big\langle[0,9],\ xy\Big\rangle,\ \Big\langle[10,12],\ \frac{3}{13}xy\Big\rangle\bigg)\ =$

\noindent$\bigg(\Big\langle[\mathtt{MININT},\leq\mathtt{MAXINT}],\ 1\Big\rangle,\ \Big\langle[0,12],\ 1\Big\rangle,\ \Big\langle[0,9],\ 0.9998500065\Big\rangle,\ \Big\langle[10,12],\ 0.23073461688\Big\rangle\bigg)$

In general, we may need infinite number of iterations to converge. This is because the height of the lattice $M$ is $\infty$. But for practical purpose we stop iterating as soon as the intervals are fixed.

With this we have extended the general theory of abstract interpretation to the unreliable program domain. But the common problem of abstract interpretation persists in this new domain as well. On of the major problems is speed -- the number of iterations required to reach a fixed program state usually depends on the numeric constants used in the program. That's why naively iterating through the collecting semantics equations over the abstract domain doesn't always scale up well. We need some additional tool to tackle this issue. In the next section we'll see why the number of iterations depends on the numeric constants used in the program and how to reduce this number to attain speed-up.

\section{Fixpoint Approximation with Convergence Acceleration by Widening}
 
\noindent Let's consider the following code segment.

\begin{lstlisting}[mathescape]
x $=_\bullet$ 0
while (x $<_\bullet$ 1000)
    x $=_\bullet$ x $+_\bullet$ 1
\end{lstlisting}

\noindent Probabilistic interval analysis will need around 1000 iterations to converge to interval $[1000, 1000]$ for $x$ at the end of the program.
This seems too slow. To reduce the number of iterations to reach a fixed point, we rely on a technique called \textit{widening}. The technique of widening basically refers to methods for accelerating the convergence of fixed point iterations. Let's first look at the formal definition of the widening operator. A widening operator $\nabla: M\times M\mapsto M$ is such that it holds the following two properties:

\begin{description}
  \item[Correctness :] \hfill
	\begin{itemize}
	  \item $\forall x,y\in M,\quad$$\GAMMA$$(x)\ \sqsubseteq_M\ $$\GAMMA$$(x\nabla y)$
	  \item $\forall x,y\in M,\quad$$\GAMMA$$(y)\ \sqsubseteq_M\ $$\GAMMA$$(x\nabla y)$
	\end{itemize}
  \item[Convergence :] \hfill
	\begin{itemize}
	  \item For all increasing chains $x^0\sqsubseteq_M x^1\sqsubseteq_M\ldots\ $, the increasing chain defined by\\
	  $y^0=x^0,\ \ldots\ ,y^{i+1}=y^i\nabla x^{i+1},\ \ldots\ $ is not strictly increasing.
	\end{itemize}
\end{description}

\subsection{Fixpoint approximation with widening}

For every \textit{collecting semantics} in abstract domain, $g^\#_j$, we write $(g^\#_j)^i$ to represent its value after $i^{th}$ iteration. Following this notation, the upward iteration sequence with widening\cite{cousot77} will be as follows:

$(g^\#_j)^0=\bot_M$

$(g^\#_j)^{i+1} =
  \begin{cases}
    (g^\#_j)^i       & \quad \text{if } sp^\#\Big((g^\#_j)^i\Big)\sqsubseteq_M (g^\#_j)^i\\
    (g^\#_j)^i\ \nabla sp^\#\Big((g^\#_j)^i\Big)  & \quad \text{otherwise}\\
  \end{cases}
$\\

\noindent This sequence is ultimately stationary and its limit, $\hat{A}$, is a sound upper approximation of \ lfp$^{^{\bot_M}}(sp^\#)$ i.e lfp$^{^{\bot_M}}(sp^\#)\sqsubseteq_M\ \hat{A}$. Now we'll use a common widening technique to implement a binary widening operator for the probabilistic interval abstract domain.

\subsection{Interval Widening with threshold set}

There are several possible schemes for implementing widening operation. We shall outline one such method as follows. A \textbf{\emph{threshold set}}, $T$ is a finite set of numbers (including $\texttt{MININT}$ and $\texttt{MAXINT}$). For a given program, we identify the set of constants used. Let this be the set $C$. We are restricting ourselves to programs with integer variables only. Now our threshold set $T=C\bigcup\left\{\texttt{MININT},\ \texttt{MAXINT}\right\}$.

Given two elements $\langle[a,b], p\rangle$ and $\langle[a',b'], p'\rangle$ in $M$, $\forall X\in M$, we define the widening with threshold operator $\nabla_T$ as follows:
\begin{definition}
  \begin{flalign*}
    X\ \nabla_T\ \bot_M&=X\\
    \bot_M\ \nabla_T\ X&=X\\
    \langle[a,b], p\rangle\ \nabla_T\ \langle[a',b'], p'\rangle&=
    \begin{cases}
      \langle[a,b], p\rangle\qquad\text{if } \langle[a',b'], p'\rangle\ \sqsubseteq_M\ \langle[a,b], p\rangle\\
      \langle[x,y], p_{xy}\rangle\qquad\text{otherwise}
    \end{cases}
  \end{flalign*}
where,

$x=\mathbf{max}\{l\in T\ |\ l\leq a'\}$,

$y=\mathbf{min}\{h\in T\ |\ h\geq b'\}$ and

$\displaystyle p_{xy}=(y-x+1)\cdot\mathbf{min}\left(\mathbf{max}\left(p.m.f\Big(\langle[a,b],p\rangle\Big),\ p.m.f\Big(\langle[a',b'],p'\rangle\Big)\right),\ \frac{1}{y-x+1}\right)$
\end{definition}

\section{Experimental Results}

\renewcommand{\arraystretch}{2.0}
\begin{table}
  \caption{Experimental Results}
  \begin{oldtabular}{|p{3.5cm}|c|p{3cm}|l|p{3cm}|}
    \hline
    \multirow{3}{*}{\centerline{Program Name}} & \multirow{3}{*}{Lines of Code} & \multicolumn{3}{c|}{Avg. time for Analysis (Number of clock ticks)\protect\footnotemark} \\
    \cline{3-5}
    & & \multirow{2}{*}{In Concrete Domain} & \multicolumn{2}{c|}{In Abstract Domain} \\
    \cline{4-5}
    & & & \multicolumn{1}{c}{Without Widening} & \multicolumn{1}{|c|}{With Widening} \\
    \hline
    Collatz Conjecture & 8 & 247.2\quad\checkmark & 54648.4\quad\checkmark & 73546.9\quad\checkmark \\
    \hline
    Factorial & 9 & 7587.3\quad\checkmark & 116.1\quad\checkmark & 31116.4\quad\checkmark \\
    \hline
    Euclid's GCD Algorithm  & 10 & 120.7\quad\checkmark & 96.4\quad\checkmark & Floating point exception (Division by 0) \\
    \hline
    Reversal of an Integer & 10 & 5475.7\quad\checkmark & 263.6\quad\checkmark & 172.4\quad\checkmark \\
    \hline
    Primality Test by Brute Force & 12 & 134.0\quad\checkmark & 102.8\quad\checkmark & 32755.7\quad\checkmark \\
    \hline
    Generate First N Fibonacci Numbers & 14 & 2556113.6 & 478.7 & 52330.1\quad\checkmark \\
    \hline
    Test Armstrong Number & 30 & 2553126.3 & 178206.3 & 1458990797.5\quad\checkmark \\
    \hline
    Generate Armstrong Numbers within a Range & 38 & Memory exhausted & 1145056.4 & 1823026061.7\quad\checkmark \\
    \hline
  \end{oldtabular}
  \label{result-table}
\end{table}
\footnotetext{Tested on an Intel Core-i5 CPU with frequency 3.2GHz.}
\renewcommand{\arraystretch}{1.0}

We have developed a prototype static analyser for \emph{C-type} programs taking hardware unreliability into consideration. It takes a program and a hardware unreliability specification as input and calculates the probabilistic intervals for each variable at each program point. As of now, our prototype is capable of parsing and analysing programs with integer variables only. It has been tested with several relatively small but popular programs. For the sake of simplicity, we had to restrict ourselves to programs that don't contain arrays, structures and pointers. The results of our testing have been shown in Table~\ref{result-table}. Not all the results converged to a fixed point. We put a limiting upper bound of 20 on the number of iterations while solving the collecting semantics iteratively. Entries marked with a tick (\checkmark) are the ones reaching fixed point within 20 iterations.

\begin{figure}
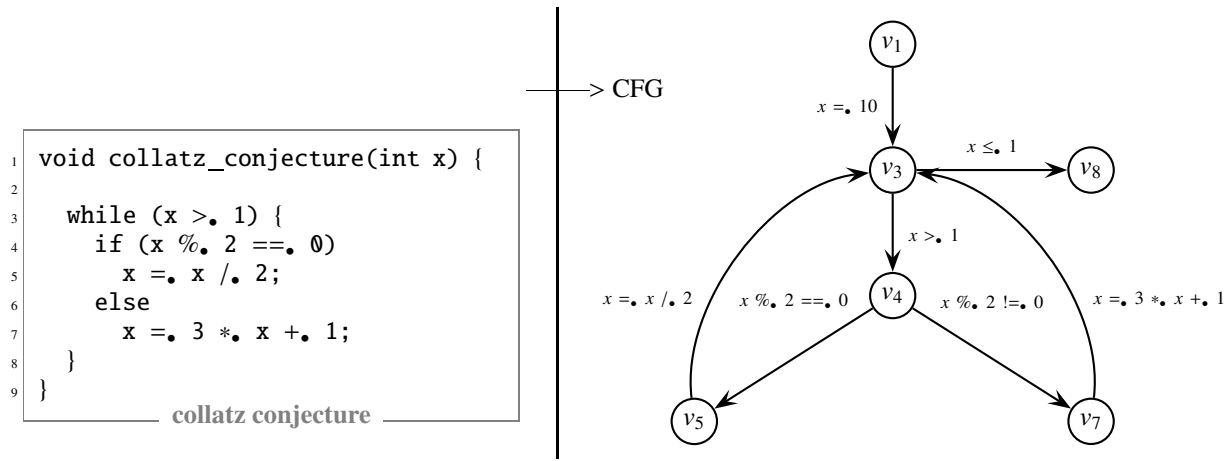

  \begin{minipage}[b]{0.40\linewidth}
    \def\snippetLabel{collatz conjecture}
    \begin{snippet}
void collatz\_conjecture(int x) \{

  while (x $>_\bullet$ 1) \{
    if (x $\%_\bullet$ 2 $==_\bullet$ 0)
      x $=_\bullet$ x $/_\bullet$ 2;
    else
      x $=_\bullet$ 3 $*_\bullet$ x $+_\bullet$ 1;
  \}
\}
    \end{snippet}
  \end{minipage}
  \rule[44mm]{4mm}{.1pt}\rule[-5mm]{1pt}{60mm}\rule[44mm]{4mm}{.1pt}\raisebox{43.2mm}{$>$ CFG}
  \begin{minipage}[b]{0.40\linewidth}
    $
    \psmatrix[colsep=2cm,rowsep=1cm,mnode=circle]
          & v_1       \\
          & v_3 & v_8 \\
          & v_4       \\
      v_5 &     & v_7
      \psset{arrowscale=2}
      \everypsbox{\scriptstyle}
      \ncline{->}{1,2}{2,2}\Bput{x\ =_\bullet\ 10}
      \ncline{->}{2,2}{3,2}>{x\ >_\bullet\ 1}
      \ncline{->}{2,2}{2,3}^{x\ \leq_\bullet\ 1}
      \ncline{->}{3,2}{4,1}^{x\ \%_\bullet\ 2\ ==_\bullet\ 0}
      \ncline{->}{3,2}{4,3}^{x\ \%_\bullet\ 2\ !=_\bullet\ 0}
      \ncarc[arcangle=45]{->}{4,1}{2,2}<{x\ =_\bullet\ x\ /_\bullet\ 2}
      \ncarc[arcangle=-45]{->}{4,3}{2,2}>{x\ =_\bullet\ 3\ *_\bullet\ x\ +_\bullet\ 1}
    \endpsmatrix
    $
  \end{minipage}
  \caption{Collatz conjecture \& its Control Flow Graph}
  \label{collatz-conjecture-figure} 
\end{figure}

Let us discuss one of the results, say that of \textit{Collatz Conjecture} in detail. Fig.~\ref{collatz-conjecture-figure} shows the code and its Control Flow Graph, which will be later used to generate the collecting semantics.

From the \textit{Control Flow Graph}, we get the \textit{Data Flow Equations} as follows:

\begin{tabular}{lcl}
$\mathtt{G}_1$ & = & $\bot$ \\
$\mathtt{G}_3$ & = & $sp(\mathtt{G}_1, x=_\bullet10)\quad\sqcup\quad sp(\mathtt{G}_5, x=_\bullet x/_\bullet2)\quad\sqcup\quad sp(\mathtt{G}_7, x=_\bullet3*_\bullet x+_\bullet1)$ \\
$\mathtt{G}_4$ & = & $sp(\mathtt{G}_3, x>_\bullet1)$ \\
$\mathtt{G}_5$ & = & $sp(\mathtt{G}_4, x\%_\bullet2==_\bullet0)$ \\
$\mathtt{G}_7$ & = & $sp(\mathtt{G}_4, x\%_\bullet2\neq_\bullet0)$ \\
$\mathtt{G}_8$ & = & $sp(\mathtt{G}_3, x\leq_\bullet1)$
\end{tabular}\\

Our tool solves these equations within finite number of iterations and generates the result shown in Table~\ref{output-table}. This result complies with the statement of the \textit{Collatz Conjecture}\cite{bendegem05}, as we can see, at the end of the program (line \#8), variable $x$ lies in the interval $[1,1]$. And in case of widening operator being used, we can confirm this interval with certainty (probability) $3.05185048982\times10^{-5}$.

\renewcommand{\arraystretch}{2}
\begin{table}
  \caption{Analysis Result for \textit{Collatz Conjecture}}
  \begin{oldtabular}{|p{1cm}|p{5cm}|c|c|}
  	\hline
  	\multicolumn{4}{|c|}{In the following results: $\qquad\qquad m=-32768\ (\mathrm{MININT})\qquad$ and $\qquad M=32767\ (\mathrm{MAXINT})$} \\
    \hline
    \multirow{2}{*}{\centerline{Line \#}} & \multirow{2}{*}{\centerline{Concrete Domain Result (8 \textit{iterations})}} & \multicolumn{2}{c|}{Abstract Domain Result} \\
    \cline{3-4}
    & & Without Widening (11 \textit{iterations}) & With Widening (4 \textit{iterations}) \\
    \hline
    \centerline{$1$} & \centerline{$x=\langle\{a\in\mathbb{Z}\ |\ m\leq a\leq M\},\ 1\rangle$} & $x=\langle[m,\ M],\ 1\rangle$ & $x=\langle[m,\ M],\ 1\rangle$ \\
    \hline
    \centerline{$3$} & \centerline{$x=\langle\{1, 2, 4, 5, 8, 10, 16\},\ 0.999996199976\rangle$} & $x=\langle[1,\ M],\ 1\rangle$ & $x=\langle[1,\ M],\ 1\rangle$ \\
    \hline
    \centerline{$4$} & \centerline{$x=\langle\{2, 4, 5, 8, 10, 16\},\ 0.999996649979\rangle$} & $x=\langle[2,\ M],\ 0.999969331495\rangle$ & $x=\langle[2,\ M],\ 1\rangle$ \\
    \hline
    \centerline{$5$} & \centerline{$x=\langle\{2, 4, 8, 10, 16\},\ 0.999996399979\rangle$} & $x=\langle[2,\ 32766],\ 0.999938563004\rangle$ & $x=\langle[2,\ 32766],\ 0.999969230565\rangle$ \\
    \hline
    \centerline{$7$} & \centerline{$x=\langle\{5\},\ 0.999998899993\rangle$} & $x=\langle[3,\ M],\ 0.999938563004\rangle$ & $x=\langle[3,\ M],\ 0.999999749998\rangle$ \\
    \hline
    \centerline{$8$} & \centerline{$x=\langle\{1\},\ 0.999996049977\rangle$} & $x=\langle[1,\ 1],\ 0.00409836004098\rangle$ & $x=\langle[1,\ 1],\ 3.05185048982\times10^{-5}\rangle$ \\
    \hline
  \end{oldtabular}
  \label{output-table}
\end{table}
\renewcommand{\arraystretch}{1.0}

\section{Applications}

We have a direct application of this analysis in some of the well known real life problems. In the following section we are going to discuss that in detail.

\subsection{Reliability Analysis of Control System Software}

\begin{figure}
  \centering
  \resizebox{\linewidth}{!}{
    \pgfdeclarelayer{foreground}
    \pgfsetlayers{main,foreground}
    \begin{tikzpicture}
      \input{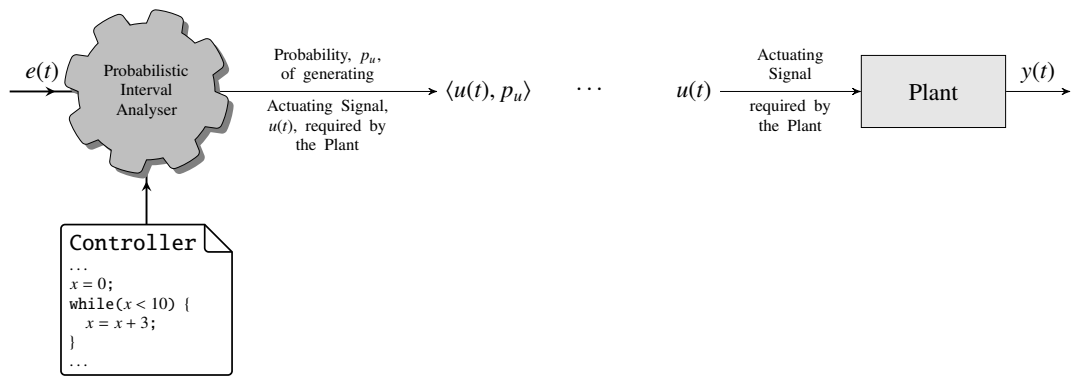}
    \end{tikzpicture}
  }
  \caption[Reliability Analysis of Control System Software]{Schematic diagram of how \emph{Probabilistic Interval Analysis} can be used for giving a reliability measurement for control system software}
  \label{reliability-analysis-application}
\end{figure}

Generally in control systems some mechanical or electrical actuators are triggered by the execution of some control program (fig.~\ref{reliability-analysis-application}). It is very common that these mechanical or electrical actuators have some predefined operating range of physical quantities like voltage, current, force etc. --- and it might malfunction or even break down if compelled to work outside that range. This physical operating range of the actuator implicitly puts a limit on the actuating signal that actuates it. These actuating signals are result of the execution of some control program. So, the program that generates these signals, is also bound by the operating range of these signals. It is actually the variables in the control program, which have predefined operating range of numerical values. If a variable is given a value from outside its operating range, it might generate a signal that could fail to actuate the actuator i.e the mechanical or electrical component. So, in the program level, it is absolutely necessary to have the knowledge of the operating range of each variable. These operating ranges of the actuating signal can be derived from the manufacturer specification of the actuator.

Now the problem is, in control systems, programs can run ceaselessly for hours or even days. Moreover they run under extreme conditions --- e.g, in an automotive control system where the temperature of engine is extremely high, some automotive control software runs for significantly long time. For this reason heating is a serious issue for control system hardware. If heating is not handled properly, which in most cases is not done, overheated hardware circuits and chips gradually become unreliable. Unreliability in hardware creeps into the software programs that run on top of it and programs start to generate unreliable results.

Here is where \emph{Probabilistic Interval Analyser} comes in. If we have the unreliability values of the hardware components at different temperatures, we can use our tool on the control system program with that unreliability and it will give a probability, $p_u$, that the desired actuating signal, $u(t)$ will be generated if the program is actually executed at that instant. So, the probability, that actuating signal is generated outside the operating range of the actuator, is $(1-p_u)$, which is the probability of \emph{actuation failure}. If we multiply $(1-p_u)$ with 100 (say), we can give an expectation of, number of actuation failures in 100 runs. This is an important metric in determining the reliability of a control system. In fig.~\ref{reliability-analysis-application}, we have shown a schematic work flow of this technique discussed above.

\section{Related Work}

Static program analysis has been a well researched area in computer science as it has major applications in wide variety of fields -- starting from compilers to as big as aviation control system and many more. All popular compilers use different techniques for statically analysing programs before or during compilation so that it can warn the end user about certain types of errors before the programs are actually executed. Static program analysis can detect errors like \emph{buffer overflow}, \emph{division by zero}, \emph{potential NULL pointer access} etc. It can also be used for code optimization like \emph{static branch prediction}, \emph{dead code elimination} etc. which is an essential part of any compiler back-end. There exist several well defined formal methods like \emph{Model Checking}, \emph{Data Flow Analysis}~\cite[pp.~35--139]{nielson99}, \emph{abstract interpretation}~\cite{cousot77} etc. which are used for static analysis of programs. Abstract interpretation is a mathematical tool for program analysis that was first introduced in 1977~\cite{cousot77}. Even since its inception, it has gained tremendous popularity among the programming language researchers' community.

There exists extremely popular and efficient static program analyser tool like ASTR{\'E}E~\cite{astree05} that uses abstract interpretation at its core. It aims at proving the absence of Run Time Errors in programs written in the C language. On personal computers, such errors, commonly found in programs, usually result in unpleasant error messages and the termination of the application, and sometimes in a system crash. In embedded applications, such errors may have graver consequences. ASTR{\'E}E analyzes structured C programs, with complex memory usages, but without dynamic memory allocation and recursion. This encompasses many embedded programs as found in earth transportation, nuclear energy, medical instrumentation, aeronautic, and aerospace applications, in particular synchronous control/command such as electric flight control~\cite{delmas07}~\cite{souyris07} or space vessels manoeuvres~\cite{bouissou09}. Apart from ASTR{\'E}E there are tools for static timing analysis also. One such tool is aiT Worst Case Execution Time (WCET) Analyzer. It computes tight upper bounds for the worst-case execution time of every possible execution scenario, with any combination of inputs. These tools use abstract interpretation as their basic building block for analysing the program. Several such tools are available at \url{https://www.absint.com}.

New abstract domains have been discovered over the years. People used the concept of abstract interpretation with \emph{interval abstract domain}~\cite{brauer10}, abstract domain for \emph{linear congruence equalities}~\cite{granger91}, abstract domains for linear combinations of program variables, such as \emph{octagon abstract domain}~\cite{octagon06}, \emph{polyhedra abstract domain}~\cite{polyhedra13} etc. All these domains are used to represent different information about the program states. There is a trade-off between the level of abstraction and the precision of results. The simpler the domain is (more abstract) the lesser information it can capture. For example the interval abstract domain is the simplest among all the aforementioned domains and it is non relational as well, i.e it yields simple intervals for individual program variables at different program points. The other abstract domains such as, linear congruence equalities domain, octagon abstract domain and polyhedra abstract domain are relational domains. They generate relational (in)equalities among two or more program variables. For a family of $n$ program variables, $x_1, x_2 \cdots x_n$, the result of analysis in linear congruence equalities domain will generate invariants of the form $\displaystyle\sum_{i=1}^n\alpha_i x_i\equiv c\ [m]$, where $\alpha_1, \alpha_2 \cdots \alpha_n, c, m$ are integers and $\equiv$ is the arithmetic congruence relation (i.e $a\equiv b\ [m]$ if and only if $\exists k\in\mathbb{Z}, a=b+km$). Octagon and polyhedra abstract domains deal with linear arithmetic constraints of general form. Octagon abstract domain is constrained to (in)equalities of at most 2 program variables, whereas polyhedra abstract domain is most general and it can handle all possible linear (in)equalities. Analysis in octagon abstract domain generates constraints of the form $\pm x_i\pm x_j\leq c$, where $x_i$ and $x_j$ range among program variables and $c$ is a constant in $\mathbb{Z}, \mathbb{Q}$ or $\mathbb{R}$. Finally the polyhedra abstract domain can generate constraints of the form $\displaystyle\sum_{i=1}^n\alpha_i x_i\leq c$, where $\alpha_1, \alpha_2 \cdots \alpha_n, c$ are constants in $\mathbb{Z}, \mathbb{Q}$ or $\mathbb{R}$. In this work, we've designed another abstract domain, the \emph{probabilistic interval abstract domain}, that is a generalization of the interval abstract domain. While the analysis in interval abstract domain yields intervals for individual program variables with a fixed probability of 1, this new non relational abstract domain relaxes that constraint and it can also generate intervals with probabilities other than 1.

Works have been done on applying abstract interpretation in the domain of probabilistic programs also. To mention a few of the notable works, we start with the work of static analysis of programs with \emph{imprecise probabilistic inputs}, which heavily relies on Dempster-Shafer structures (DSI) or P-boxes~\cite{vstte14}. This analyser assumes each input value to be lying within a \emph{confidence box} $[\underline{P},\overline{P}]$. It analyses programs using random generators or random inputs and also allow non-deterministic inputs, not necessarily following a random distribution. Based on this assumption, it works on to perform the interval analysis of each program variable. Few other works are there, where people statically analyse programs with non-deterministic and probabilistic behaviour~\cite{monn00}~\cite{monn01}~\cite{sumit13} using general abstract interpretation based method.

Our work is orthogonal to all these as we are concerned with the imprecision seeded into the hardware rather than in input. So, in our case inputs come from reliable sources with reliable values. Essentially we'll be designing a new abstract domain that takes care of the hardware unreliability, during the course of this work.

\section{Conclusion}

Result obtained by simple interval analysis is not much precise (even though it is safe), especially when widening is used. There are other abstract domains that can give more precise results. People have already designed abstract domains like \emph{octagon abstract domain}~\cite{octagon06}, \emph{polyhedra abstract domain}~\cite{polyhedra13} etc. which are little complex than the simple \emph{interval abstract domain} but they generate much precise results. Unlike interval abstract domain, which treats each program variable separately, octagon abstract domain works on a linear combination of two variables and polyhedra abstract domain works on a linear combination of arbitrary number of variables at a time. The measure of imprecision in the result decreases as we move from interval to octagon to polyhedra abstract domain respectively. Scope is there for extending these two abstract domains, octagon and polyhedra, for unreliable programs as well. We can expect them to improve the result of the analysis here as well, taking care of the hardware unreliability at the same time.


\newcommand{\etalchar}[1]{$^{#1}$}

\label{lastpage}


\begin{thebibliography}{ABG{\etalchar{+}}14}

\bibitem[ABG{\etalchar{+}}14]{vstte14}
  Adje, A., Bouissou, O., Goubault-Larrecq, J., Goubault, E. and Putot, S.:
  \newblock {\em Verified Software: Theories, Tools, Experiments: $5^{th}$ International Conference, VSTTE 2013, Menlo Park, CA, USA, May 17-19, 2013, Revised Selected Papers}, ch. Static Analysis of Programs with Imprecise Probabilistic Inputs, pp. 22--47.
  \newblock Springer Berlin Heidelberg, 2014.
  \newblock \url{http://dx.doi.org/10.1007/978-3-642-54108-7_2}.

\bibitem[Ben05]{bendegem05}
  Bendegem, J. V.:
  \newblock The collatz conjecture. a case study in mathematical problem solving.
  \newblock {\em Logic and Logical Philosophy}, 14(1):7--23, 2005.
  \newblock \url{http://dx.doi.org/10.12775/LLP.2005.002}.

\bibitem[BNS10]{brauer10}
  Brauer, J., Noll, T., and Schlich, B.:
  \newblock Interval analysis of microcontroller code using abstract interpretation of hardware and software.
  \newblock In {\em Proceedings of the $13^{th}$ International Workshop on Software \&\#38; Compilers for Embedded Systems}, pp. 3:1--3:10, 2010.
  \newblock \url{http://doi.acm.org/10.1145/1811212.1811216}.

\bibitem[CoC77]{cousot77}
  Cousot, P. and Cousot, R.:
  \newblock Abstract interpretation: A unified lattice model for static analysis of programs by construction or approximation of fixpoints.
  \newblock In {\em Proceedings of the $4^{th}$ ACM SIGACT-SIGPLAN Symposium on Principles of Programming Languages}, pp. 238--252, 1977.
  \newblock \url{http://doi.acm.org/10.1145/512950.512973}.

\bibitem[CCF{\etalchar{+}}05]{astree05}
  Cousot, P., Cousot, R., Feret, J., Mauborgne, L., Min{\'e}, A., Monniaux, D. and Rival, X.:
  \newblock {\em Programming Languages and Systems: $14^{th}$ European Symposium on Programming, ESOP 2005, Held as Part of the Joint European Conferences on Theory and Practice of Software, ETAPS 2005, Edinburgh, UK, April 4-8, 2005. Proceedings}, ch. The ASTRE{\'E} Analyzer, pp. 21--30.
  \newblock Springer Berlin Heidelberg, 2005.
  \newblock \url{http://dx.doi.org/10.1007/978-3-540-31987-0_3}.

\bibitem[CGM{\etalchar{+}}08]{chakrapani08}
  Chakrapani, L. N. B., George, J., Marr, B., Akgul, B. E. S. and Palem, K. V.:
  \newblock {\em VLSI-SoC: Research Trends in VLSI and Systems on Chip: Fourteenth International Conference on Very Large Scale Integration of System on Chip (VLSI-SoC2006), October 16-18, 2006, Nice, France}, ch. Probabilistic Design: A Survey of Probabilistic CMOS Technology and Future Directions for Terascale IC Design, pp. 101--118.
  \newblock Springer US, 2008.
  \newblock \url{http://dx.doi.org/10.1007/978-0-387-74909-9_7}.

\bibitem[CMR13]{carbin13}
  Carbin, M., Misailovic, S. and Rinard, M. C.:
  \newblock Verifying quantitative reliability for programs that execute on unreliable hardware.
  \newblock {\em SIGPLAN Not.}, 48(10), 2013.
  \newblock \url{http://doi.acm.org/10.1145/2544173.2509546}.

\bibitem[Esc69]{rascel69}
  Esch, J. W.:
  \newblock {\em Rascel, a Programmable Analog Computer Based on a Regular Array of Stochastic Computing Element Logic}.
  \newblock PhD thesis, University of Illinois at Urbana–Champaign, Champaign, IL, USA, 1969.
  \newblock \url{https://archive.org/details/rascelprogrammab332esch}.

\bibitem[Gra91]{granger91}
  Granger, P.:
  \newblock {\em TAPSOFT '91: Proceedings of the International Joint Conference on Theory and Practice of Software Development Brighton, UK, April 8-12, 1991}, ch. Static analysis of linear congruence equalities among variables of a program, pp. 169--192.
  \newblock Springer Berlin Heidelberg, 1991.
  \newblock \url{http://dx.doi.org/10.1007/3-540-53982-4_10}.

\bibitem[Har77]{harrison77}
  Harrison, W. H.:
  \newblock Compiler analysis of the value ranges for variables.
  \newblock {\em IEEE Transactions on Software Engineering}, SE-3(3):243--250, 1977.
  \newblock \url{https://dx.doi.org/10.1109/TSE.1977.231133}.

\bibitem[Min06]{octagon06}
  Min{\'e}, A.:
  \newblock The octagon abstract domain.
  \newblock {\em Higher-Order and Symbolic Computation}, 19(1):31--100, 2006.
  \newblock \url{http://dx.doi.org/10.1007/s10990-006-8609-1}.

\bibitem[Mon00]{monn00}
  Monniaux, D.:
  \newblock {\em Static Analysis: $7^{th}$ International Symposium, SAS 2000, Santa Barbara, CA, USA, June 29 - July 1, 2000. Proceedings}, chapter Abstract Interpretation of Probabilistic Semantics, pp. 322--339.
  \newblock Springer Berlin Heidelberg, 2000.
  \newblock \url{http://dx.doi.org/10.1007/978-3-540-45099-3_17}.

\bibitem[Mon01]{monn01}
  Monniaux, D.:
  \newblock {\em Programming Languages and Systems: $10^{th}$ European Symposium on Programming, ESOP 2001 Held as Part of the Joint European Conferences on Theory and Practice of Software, ETAPS 2001 Genova, Italy, April 2-6, 2001 Proceedings}, ch. Backwards Abstract Interpretation of Probabilistic Programs, pp. 367--382.
  \newblock Springer Berlin Heidelberg, 2001.
  \newblock \url{http://dx.doi.org/10.1007/3-540-45309-1_24}.

\bibitem[MWG05]{marpe05}
  Marpe, D., Wiegand, T. and Gordon, S.:
  \newblock H.264/mpeg4-avc fidelity range extensions: tools, profiles, performance, and application areas.
  \newblock In {\em Image Processing, 2005. ICIP 2005. IEEE International Conference on}, pages I--593--6, 2005.
  \newblock \url{http://dx.doi.org/10.1109/ICIP.2005.1529820}.

\bibitem[NNH99]{nielson99}
  Nielson, F., Nielson, H. R. and Hankin, C.:
  \newblock {\em Principles of Program Analysis}.
  \newblock Springer-Verlag Berlin Heidelberg, 1999.
  \newblock \url{https://dx.doi.org/10.1007/978-3-662-03811-6}.

\bibitem[SeB13]{polyhedra13}
  Seladji, Y. and Bouissou, O.:
  \newblock {\em Verification, Model Checking, and Abstract Interpretation: $14^{th}$ International Conference, VMCAI 2013, Rome, Italy, January 20-22, 2013. Proceedings}, ch. Fixpoint Computation in the Polyhedra Abstract Domain Using Convex and Numerical Analysis Tools, pp. 149--168.
  \newblock Springer Berlin Heidelberg, 2013.
  \newblock \url{http://dx.doi.org/10.1007/978-3-642-35873-9_11}.

\bibitem[SCG13]{sumit13}
  Sankaranarayanan, S., Chakarov, A. and Gulwani, S.:
  \newblock Static analysis for probabilistic programs: inferring whole program properties from finitely many paths.
  \newblock {\em SIGPLAN Not.}, pp. 447--458, 2013.
  \newblock \url{http://doi.acm.org/10.1145/2499370.2462179}.

\bibitem[DeS07]{delmas07}
  Delmas, D. and Souyris, J.:
  \newblock {\em Static Analysis: $14^{th}$ International Symposium, SAS 2007, Kongens Lyngby, Denmark, August 22-24, 2007. Proceedings}, ch. Astr{\'e}e: From Research to Industry, pp. 437--451.
  \newblock Springer Berlin Heidelberg, 2007.
  \newblock \url{http://dx.doi.org/10.1007/978-3-540-74061-2_27}.
  
\bibitem[SoD07]{souyris07}
  Souyris, J. and Delmas, D.:
  \newblock {\em Computer Safety, Reliability, and Security: $26^{th}$ International Conference, SAFECOMP 2007, Nuremberg, Germany, September 18-21, 2007. Proceedings}, ch. Experimental Assessment of Astr{\'e}e on Safety-critical Avionics Software, pp. 479--490.
  \newblock Springer Berlin Heidelberg, 2007.
  \newblock \url{http://dx.doi.org/10.1007/978-3-540-75101-4_45}.
  
\bibitem[BCC{\etalchar{+}}08]{bouissou09}
  Bouissou, O., Conquet, E., Cousot, P., Cousot, R., Feret, J., Ghorbal, K., Goubault, E., Lesens, D., Mauborgne, L., Min{\'e}, A., Putot, S., Rival, X. and Turin, M.:
  \newblock {\em International Space System Engineering Conference, $13^{th}$ Data Systems In Aerospace, DASIA'09, Istanbul, Turkey, May 26--29, 2009, Proceedings}, ch. Space software validation using abstract interpretation, pp. 1--7.
  \newblock Eurospace, Paris, 2007.


\end{thebibliography}
\end{document}